\documentclass[useAMS,usenatbib]{mn2e}
\usepackage{graphicx,txfonts}
\def \aap{A\&A}
\def \aaps{A\&AS}
\def \al{Astron.~Lett.}
\def \apj{ApJ}
\def \apjl{ApJ}
\def \cjc{Canadian~J.~Chem.}
\def \cjp{Canadian~J.~Phys.}
\def \epjc{Euro.~Phys.~J.~C}
\def \epjst{Euro.~Phys.~J.~Special~Topics}
\def \jms{J.~Molecular~Spectrosc.}
\def \josb{J.~Opt.~Soc.~America~B}
\def \lnp{Lecture~Notes~Phys.}
\def \jpb{J.~Phys.~B}
\def \met{Metrologia}
\def \mnras{MNRAS}
\def \mp{Mol.~Phys.}
\def \mpla{Mod.~Phys.~Lett.~A}
\def \prd{Phys.~Rev.~D}
\def \prl{Phys.~Rev.~Lett.}
\def \sci{Sci}
\def \spjetpl{Sov.~Phys.~JETP~Lett.}

\usepackage{color}

\newcommand{\ms}{\hbox{${\rm m\,s}^{-1}$}}
\newcommand{\kms}{\hbox{${\rm km\,s}^{-1}$}}

\newcommand{\SNR}{\hbox{${\rm SNR}$}}
\newcommand{\dmu}{\hbox{$\Delta\mu/\mu$}}

\newcommand{\bspsmall}{\vspace{0.5cm}\small\noindent This paper has been typeset
from a \TeX/\LaTeX\ file prepared by the author.\normalsize}

\title[Constraint on the variation of $\mu$]{Keck telescope constraint
  on cosmological variation of the proton-to-electron mass ratio}

\author[A. L. Malec et al.]{A. L. Malec,$^{1}$\thanks{E-mail:
    amalec@swin.edu.au (ALM)} R. Buning,$^{2}$ M. T. Murphy,$^{1}$ N.
  Milutinovic,$^{3}$ S. L. Ellison,$^{3}$\newauthor J. X. Prochaska,$^{4}$ L.
  Kaper,$^{2,5}$ J. Tumlinson,$^{6}$ R. F. Carswell,$^{7}$ W. Ubachs$^{2}$\\
  $^{1}$Centre for Astrophysics and Supercomputing, Swinburne University of Technology, Melbourne, Victoria 3122, Australia\\
  $^{2}$Laser Centre, VU University, De Boelelaan 1081, 1081 HV Amsterdam, The Netherlands\\
  $^{3}$Department of Physics and Astronomy, University of Victoria, Victoria, BC, V8P 1A1, Canada\\
  $^{4}$University of California Observatories -- Lick Observatory, University of California, Santa Cruz, CA 95064\\
  $^{5}$Astronomical Institute Anton Pannekoek, Universiteit van Amsterdam, 1098 SJ Amsterdam, The Netherlands\\
  $^{6}$Yale Center for Astronomy and Astrophysics, Department of Physics, New Haven, CT 06520, USA\\
  $^{7}$Institute of Astronomy, University of Cambridge, Madingley Road, Cambridge, CB3 0HA, UK
}

\begin{document}

\date{Accepted 2009 December 14.  Received 2009 November 18; in
  original form 2009 June 9}

\pagerange{\pageref{firstpage}--\pageref{lastpage}}

\pubyear{2010}

\maketitle

\label{firstpage}

\begin{abstract}
  Molecular transitions recently discovered at redshift $z_{\rm
    abs}=2.059$ toward the bright background quasar J2123$-$0050 are
  analysed to limit cosmological variation in the proton-to-electron
  mass ratio, $\mu\equiv m_{\rm p}/m_{\rm e}$. Observed with the Keck
  telescope, the optical echelle spectrum has the highest resolving
  power and largest number (86) of H$_2$ transitions in such analyses
  so far.  Also, (seven) HD transitions are used for the first time to
  constrain $\mu$-variation. These factors, and an analysis employing
  the fewest possible free parameters, strongly constrain $\mu$'s
  relative deviation from the current laboratory value: $\dmu
  =(+5.6\pm5.5_{\rm stat}\pm2.9_{\rm sys})\times10^{-6}$, indicating
  an insignificantly larger $\mu$ in the absorber. This is the first
  Keck result to complement recent null constraints from three systems
  at $z_{\rm abs}>2.5$ observed with the Very Large Telescope.  The
  main possible systematic errors stem from wavelength calibration
  uncertainties. In particular, distortions in the wavelength solution
  on echelle order scales are estimated to contribute approximately
  half the total systematic error component, but our estimate is model
  dependent and may therefore under or overestimate the real effect,
  if present.

  To assist future $\mu$-variation analyses of this kind, and other
  astrophysical studies of H$_2$ in general, we provide a compilation
  of the most precise laboratory wavelengths and calculated parameters
  important for absorption-line work with H$_2$ transitions redwards of
  the hydrogen Lyman limit.
\end{abstract}


\section{Introduction}\label{sec:intro}

The Standard Model of particle physics is parametrized by several
dimensionless `fundamental constants', such as coupling constants and
mass ratios, whose values are not predicted by the Model itself.
Instead their values and, indeed, their constancy must be established
experimentally. If found to vary in time or space, understanding their
dynamics may require a more fundamental theory, perhaps one unifying
the four known physical interactions. Two such parameters whose
constancy can be tested to high precision are the fine-structure
constant, $\alpha\equiv e^2/4\pi\epsilon_0 \hbar c$, characterising
electromagnetism's strength, and the proton-to-electron mass ratio,
$\mu\equiv m_{\rm p}/m_{\rm e}$ -- effectively the ratio of the strong
and electro-weak scales. Grand Unified Theories can predict
relationships between variations in $\mu$ and $\alpha$
\citep[e.g.][]{CalmetX_02a}, but which parameter varies the most and
whether in the same or opposite sense to the other is model-dependent
\citep{DentT_08a}.

Earth-bound laboratory experiments, conducted over several-year
time-scales, which use ultra-stable lasers to compare different atomic
clocks based on different atoms/ions \citep[e.g.~Cs, Hg$^+$, Al$^+$,
Yb$^+$, Sr, Dy;
e.g.][]{PrestageJ_95a,MarionH_03a,PeikE_04a,CingozA_07a}, have limited
$\alpha$'s time-derivative to
$\dot{\alpha}/\alpha=(-1.6\pm2.3)\times10^{-17}{\rm \,yr}^{-1}$
\citep{RosenbandT_08a}. Combining results from similar experiments
limits $\mu$'s time-derivative to
$\dot{\mu}/\mu=(-1.6\pm1.7)\times10^{-15}{\rm \,yr}^{-1}$
\citep{BlattS_08a}. A direct constraint was also recently obtained
from a ro-vibrational transition in the SF$_6$ molecule,
$\dot{\mu}/\mu=(-3.8\pm5.6)\times10^{-14}{\rm \,yr}^{-1}$
\citep{ShelkovnikovA_08a}.

Important probes of variations over much larger space- and time-scales
-- up to $\sim$90\% of the age of the Universe -- are narrow
absorption lines imprinted on the spectra of distant, background
quasars by gas clouds associated with foreground galaxies
\citep{BahcallJ_67b}. In particular, the Lyman and Werner transitions
of molecular hydrogen (laboratory wavelengths $\lambda_{\rm
  lab}\la1150$\,\AA), by far the most abundant molecule in the
Universe, are useful $\mu$-variation indicators at cosmological
redshifts $z>2$ where they are detectable with ground-based telescopes
\citep{ThompsonR_75a,VarshalovichD_93b}.

Variations in $\mu$ should shift the ro-vibronic transition
frequencies in molecular spectra. This mass-dependent shift is
quantified by a sensitivity coefficient, $K_i$, for each transition
$i$. Consider a single transition arising in an absorption cloud whose
redshift is established to be $z_{\rm abs}$ from other transitions
which are insensitive to variations in $\mu$. If $\mu$ was the same in
the absorption cloud as in the laboratory, we would expect to find
transition $i$ at wavelength $\lambda_{\rm lab}^i(1+z_{\rm
  abs})$. If instead we measure it to be at wavelength
$\lambda_i=\lambda_{\rm lab}^i(1+z_i)$ (i.e.~at redshift $z_i\ne
z_{\rm abs}$), then the shift in redshift, $\Delta z_i\equiv z_i -
z_{\rm abs}$, or velocity, $\Delta v_i$, can be ascribed to a
variation in $\mu$,
\begin{equation}
  \frac{\Delta v_i}{c} \approx \frac{\Delta z_i}{1+z_{\rm abs}} = K_i \frac{\Delta\mu}{\mu}\,,
\label{eq:shifts}
\end{equation}
where $\Delta\mu/\mu\equiv(\mu_z-\mu_{\rm lab})/\mu_{\rm lab}$ for
$\mu_{\rm lab}$ and $\mu_z$ the current laboratory value of $\mu$ and
its value in the absorption cloud at redshift $z$, respectively. That
different transitions have different $K$ values enables a
differential measurement of $\Delta\mu/\mu$ from two or more
transitions. That is, it allows the redshift $z_{\rm abs}$ to be
determined simultaneously with $\Delta\mu/\mu$.

Indications for a significantly positive $\Delta\mu/\mu$ have been
derived from two H$_2$-bearing quasar absorbers observed with the
Ultraviolet and Visual Echelle Spectrograph (UVES) on the ESO Kueyen
Very Large Telescope (VLT) in Chile. \citet{IvanchikA_05a} studied
high resolving power ($R\approx53000$), high signal-to-noise ratio
($\SNR\approx30$--$70$) UVES spectra of the $z_{\rm abs}=2.595$ and
$3.025$ absorbers towards the quasars Q\,0405$-$443 and Q\,0347$-$383,
respectively. They used a total of 76 H$_2$ lines in the two spectra,
fitting them `line-by-line' -- i.e.~independently of each other -- to
derive values for $\lambda_i$. In this analysis the H$_2$ absorption
profiles in Q\,0347$-$383 were treated as comprising a single cloud.
Of the two resolved H$_2$ features observed in Q\,0405$-$443, only the
strongest was fitted. Using two different sets of H$_2$ laboratory
wavelengths (\citealt{AbgrallH_93c} and \citealt{PhilipJ_04a})
\citeauthor{IvanchikA_05a} derived two sets of $z_i$ values which
yielded two $\Delta\mu/\mu$ values, $(+30.5\pm7.5)\times10^{-6}$ and
$(+16.5\pm7.4)\times10^{-6}$ respectively.

\citet{ReinholdE_06a} subsequently improved the laboratory wavelengths
and the calculation of the $K$ sensitivity coefficients (see
Sections \ref{ssec:lab} and \ref{ssec:K}). Using the same values of
$\lambda_i$ and their uncertainties derived by \citet{IvanchikA_05a},
\citeauthor{ReinholdE_06a} performed a `line-by-line' analysis to
refine the values of $\Delta\mu/\mu$ for each absorber:
$(+20.6\pm7.9)\times10^{-6}$ for Q\,0347$-$383 and
$(+27.8\pm8.8)\times10^{-6}$ for Q\,0405$-$443. The combined value of
$(+24.5\pm5.9)\times10^{-6}$ was presented as an \emph{indication} for
cosmological variation in $\mu$.

\citet{KingJ_08a} reanalysed the same raw UVES quasar spectra with
improved flux extraction and, more importantly, using the improved
wavelength calibration procedures detailed in \citet{MurphyM_07b}.
\citeauthor{KingJ_08a} used slightly more H$_2$ transitions and a
`simultaneous fitting' technique with explicit treatment of
Lyman-$\alpha$ forest lines (see Section \ref{ssec:chi}) to constrain
the values of $\Delta\mu/\mu$ in the newly reduced and calibrated
spectra. They found decreased values in both absorbers compared to
previous works: $\Delta\mu/\mu=(+8.2\pm7.4)\times10^{-6}$ and
$(+10.1\pm6.2)\times10^{-6}$ for Q\,0347$-$383 and Q\,0405$-$443
respectively.  \citeauthor{KingJ_08a} also analysed 64 H$_2$ lines of
a third absorption system, that at $z_{\rm abs}=2.811$ towards
Q\,0528$-$250, using UVES spectra with $R\approx45000$ and
$\SNR\approx25$--$45$. This provided the tightest constraint of all
three absorbers: $\Delta\mu/\mu=(-1.4\pm3.9)\times10^{-6}$. Thus, the
combined result, where the slightly positive values for Q\,0347$-$383
and Q\,0405$-$443 are somewhat cancelled by the slightly negative but
more precise value for Q\,0528$-$250, was a null constraint of
$\Delta\mu/\mu=(+2.6\pm3.0)\times10^{-6}$. The same UVES spectra were
also recently studied by \citet{WendtM_08a} and \citet{ThompsonR_09a}
using different data reduction and analysis techniques, generally
aimed at avoiding and/or understanding potential systematic errors and
biases in $\Delta\mu/\mu$. They also find null constraints, albeit
with somewhat larger statistical errors than \citeauthor{KingJ_08a}
due to their more conservative approaches.

So, while some analyses indicate a varying $\mu$
\citep[e.g.][]{ReinholdE_06a}, recent reanalyses of the three
well-documented and high-quality H$_2$ absorption spectra at $z>2$ do
not provide evidence for cosmological variation in $\mu$. A much
larger statistical sample is desirable, deriving from several
telescopes and spectrographs, for definitive results and to provide
measurements over a larger redshift range. However, the scarcity of
known H$_2$ absorbers hampers such progress: $>$1000 absorption
systems rich in neutral hydrogen -- i.e.~damped Lyman-$\alpha$
systems, with H{\sc \,i} column densities $N($H{\sc
  \,i}$)\ge2\times10^{20}$\,cm$^{-2}$ -- are known
\citep[e.g.][]{ProchaskaJ_09a} but systematic searches for H$_2$ have
been conducted in $<$100 systems \citep[e.g.][]{LedouxC_03a} and only
$\sim$15 are known to harbour detectable column densities of H$_2$
\citep{NoterdaemeP_08a}. Furthermore, for varying-$\mu$ analyses,
H$_2$ absorbers must be (i) at $z>2$ to shift enough Lyman and Werner
transitions above the atmospheric cutoff ($\sim$3000\,\AA), (ii) have
bright background quasars to enable high-\SNR, high resolution
spectroscopy and, (iii) have high enough H$_2$ column densities so
that individual transitions absorb significant fractions of the quasar
continuum.

In these respects, and others, the newly discovered absorber studied
here -- at $z_{\rm abs}=2.059$ towards the $z_{\rm em}=2.261$ quasar
SDSS J212329.46$-$005052.9 (hereafter J2123$-$0050; Milutinovic et
al.,~in preparation) -- is exceptional. It is the first studied with
the High Resolution Echelle Spectrometer \citep[HIRES;][]{VogtS_94a}
on the Keck telescope in Hawaii to provide constraints of similar
precision to the three VLT ones\footnote{\citet{CowieL_95a} previously
  analysed Keck/HIRES spectra of Q\,0528$-$250 to derive a
  comparatively weak constraint of
  $\Delta\mu/\mu=(-80\pm313)\times10^{-6}$.}. J2123$-$0050 is
unusually bright ($r$-band magnitude $\approx$\,16.5\,mag), providing
a high \SNR\ spectrum in just 6 hours observation with, importantly,
the highest spectral resolution for an H$_2$ absorber to date:
$R\approx110000$ or full-width-at-half-maximum ${\rm FWHM}\approx
2.7$\,\kms. HIRES's high ultraviolet (UV) throughput provides 86 H$_2$
transitions for constraining $\mu$-variation, the largest number in an
individual absorber so far. Also, for the first time, (7) HD
transitions are used to constrain $\mu$-variation; HD has been
observed in just two other high-redshift absorbers
\citep{VarshalovichD_01a,NoterdaemeP_08b}.

This paper is organised as follows. The data required for our analysis
-- the Keck spectrum of J2123$-$0050, laboratory H$_2$/HD transition
wavelengths and the sensitivity coefficients -- are described in
Section \ref{sec:data}. Section \ref{sec:analysis} presents our
analysis technique and main result while in Section \ref{sec:sys} we
conduct a variety of internal consistency checks and consider the most
important systematic errors. We conclude with a comparison of our
results with others in the literature in Section \ref{sec:discussion}.

\section{Data}\label{sec:data}

\subsection{Keck spectrum of J2123$-$0050}\label{ssec:keck}

\begin{figure*}
\begin{center}
\includegraphics[width=0.92\textwidth]{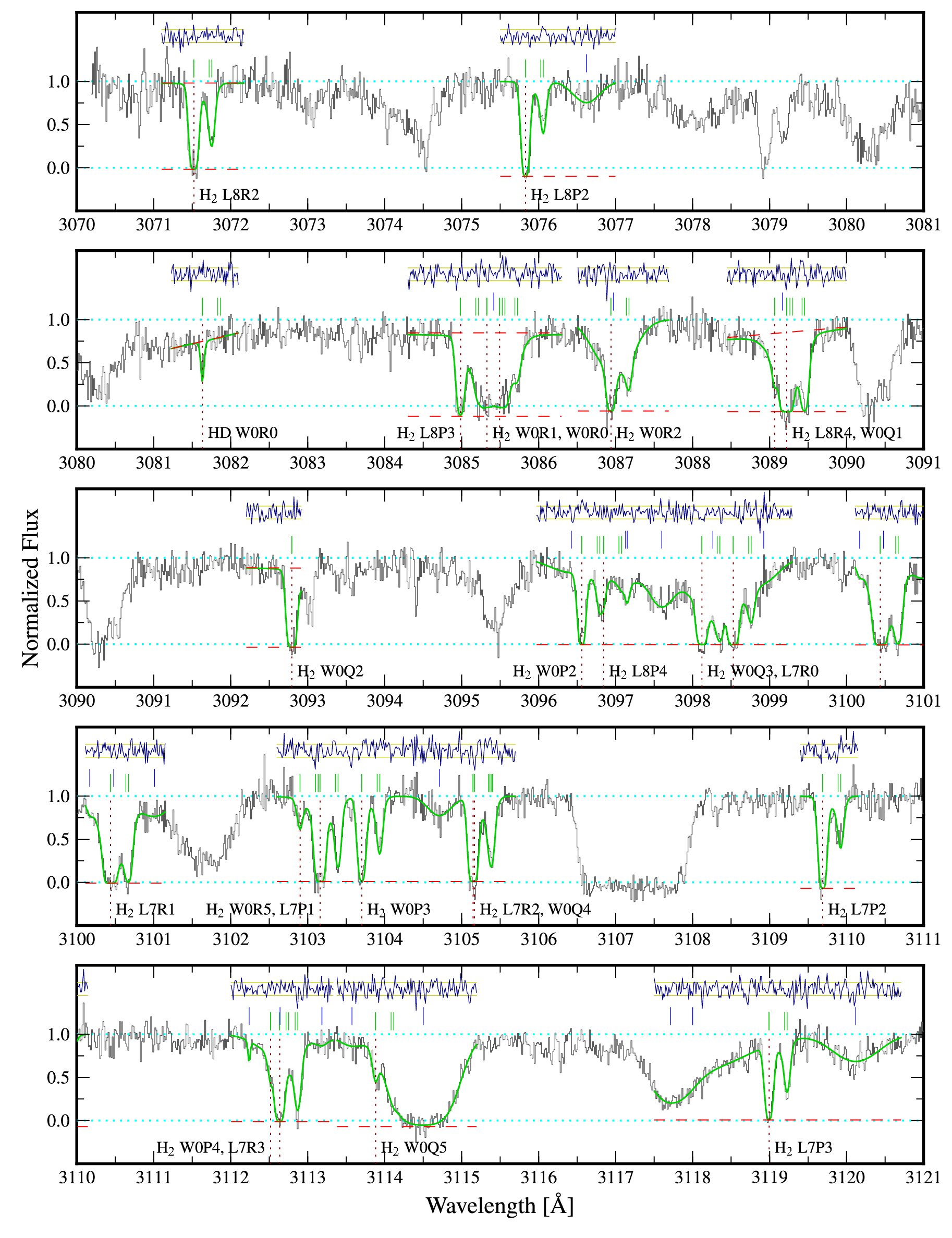}\vspace{-0.5em}
\caption{All regions of the J2123$-$0050 Keck spectrum fitted
  simultaneously in our analysis. Here we show only 5 of the 37 panels
  comprising the full figure; the other panels are available in the
  electronic version of this paper. The spectrum (black histogram) is
  normalized by a nominal continuum (upper dotted line) fitted over
  large spectral scales. Local linear continua (upper dashed lines)
  and zero levels (lower dashed lines) are fitted simultaneously with
  the H$_2$/HD and broader Lyman-$\alpha$ lines.  The fits are shown
  with solid grey/green lines. H$_2$/HD transitions are labelled and
  their constituent velocity components are indicated by grey/green
  tick-marks immediately above the spectrum. Higher above the spectrum
  are tick-marks indicating the positions of Lyman-$\alpha$ lines
  (dark-grey/blue) and Fe{\sc \,ii} lines (lighter grey/red). Note
  that the metal-line velocity structure is constrained with the
  Fe{\sc \,ii}\,$\lambda$1608\,\AA\ transition shown in the final
  panel of the figure (i.e.~in the electronic version). The residual
  spectrum (i.e.~$[{\rm data}] - [{\rm fit}]$), normalized to the
  1-$\sigma$ errors (faint, horizontal solid lines), is shown above
  the tick-marks.}
\label{fig:fit_all}
\end{center}
\end{figure*}

Full details of the Keck/HIRES observations of J2123$-$0050 are
provided in Milutinovic et al.~(in preparation) who study the physical
conditions and elemental abundances in the $z_{\rm abs}=2.059$
absorber. Here we present only the features of the spectral data
reduction and the final spectrum which are important for the
varying-$\mu$ analysis.

The HIRES observations comprised 2 sets of 3$\times$1-hr exposures of
J2123$-$0050 taken on consecutive nights (2007 August 19 \& 20). The
seeing was very good, just 0\farcs3--0\farcs5 at $\sim$6000\,\AA\
(where the guide camera is sensitive), and so a 0\farcs4-wide slit was
used. A single thorium--argon (ThAr) lamp exposure calibrated each set
of three quasar exposures; the ThAr exposure was taken immediately
before or after each quasar set. Spectrograph temperature and
atmospheric pressure shifts between the quasar and ThAr exposures were
$<$1\,K and $<1$\,mbar respectively.

The raw spectra were reduced using the {\sc hiredux}
software\footnote{http://www.ucolick.org/$\sim$xavier/HIRedux} written
and maintained by one of us (JXP). The flux extraction procedure also
calculates the formal statistical flux error spectrum for each echelle
order.  Particular attention was paid to accurate wavelength
calibration. ThAr lines were pre-selected with the procedures
described in \citet{MurphyM_07b} and improvements were made to the
ThAr line fitting in {\sc hiredux} to ensure that reliable centroids
for the ThAr lines were determined. The ThAr flux was extracted using
the same spatial profile weights as the corresponding quasar
extraction. The root-mean-square (RMS) of the wavelength calibration
residuals was $\sim$80\,\ms; that is, the calibration at any given
wavelength has RMS error $\sim$80\,\ms. The resolving power was
measured from the extracted ThAr spectra to be $R\approx110000$.
During extraction, the spectra were re-binned to a common
vacuum-heliocentric wavelength scale with a dispersion of
1.3\,\kms\,pixel$^{-1}$. They were combined to form a final spectrum
for subsequent analysis using the {\sc uves\_popler}
software\footnote{http://astronomy.swin.edu.au/$\sim$mmurphy/UVES\_popler.html}
written and maintained by one of us (MTM).

The final spectrum of J2123$-$0050 covers (vacuum--heliocentric)
wavelengths 3071--5896\,\AA. An initial global quasar continuum was
constructed using simple polynomial fits. The details of this process
are unimportant because the local continuum was modelled and fitted
simultaneously with most H$_2$/HD and Lyman-$\alpha$ forest absorption
lines. That is, the initial global continuum is just a nominal, fixed
starting guess for most molecular transitions. All the H$_2$/HD
transitions fall bluewards of 3421\,\AA\ where the \SNR\ in the
nominal continuum ranged from 7 per 1.3\,\kms\ pixel at
$\sim$3075\,\AA\ to 25 at $\sim$3420\,\AA. We made a final, manual
check on the data quality in the spectral region around each H$_2$/HD
transition. In particular, we ensured that the data contributed by the
different exposures agreed to within the uncertainty expected from the
individual flux error arrays derived during the extraction procedure.
This is also a check that the statistical uncertainties fairly
estimate the RMS flux variations in the final spectrum. Also, no
significant wavelength or velocity shifts between the individual
spectra were apparent.

Figure \ref{fig:fit_all} shows the region of the final spectrum
containing H$_2$/HD lines (3071--3421\,\AA) and a region redwards of
the Lyman-$\alpha$ emission line (4905--4926\,\AA) containing metal
lines which are required in our fit described in Section
\ref{ssec:chi}.

\begin{figure}
\begin{center}
\includegraphics[width=\columnwidth]{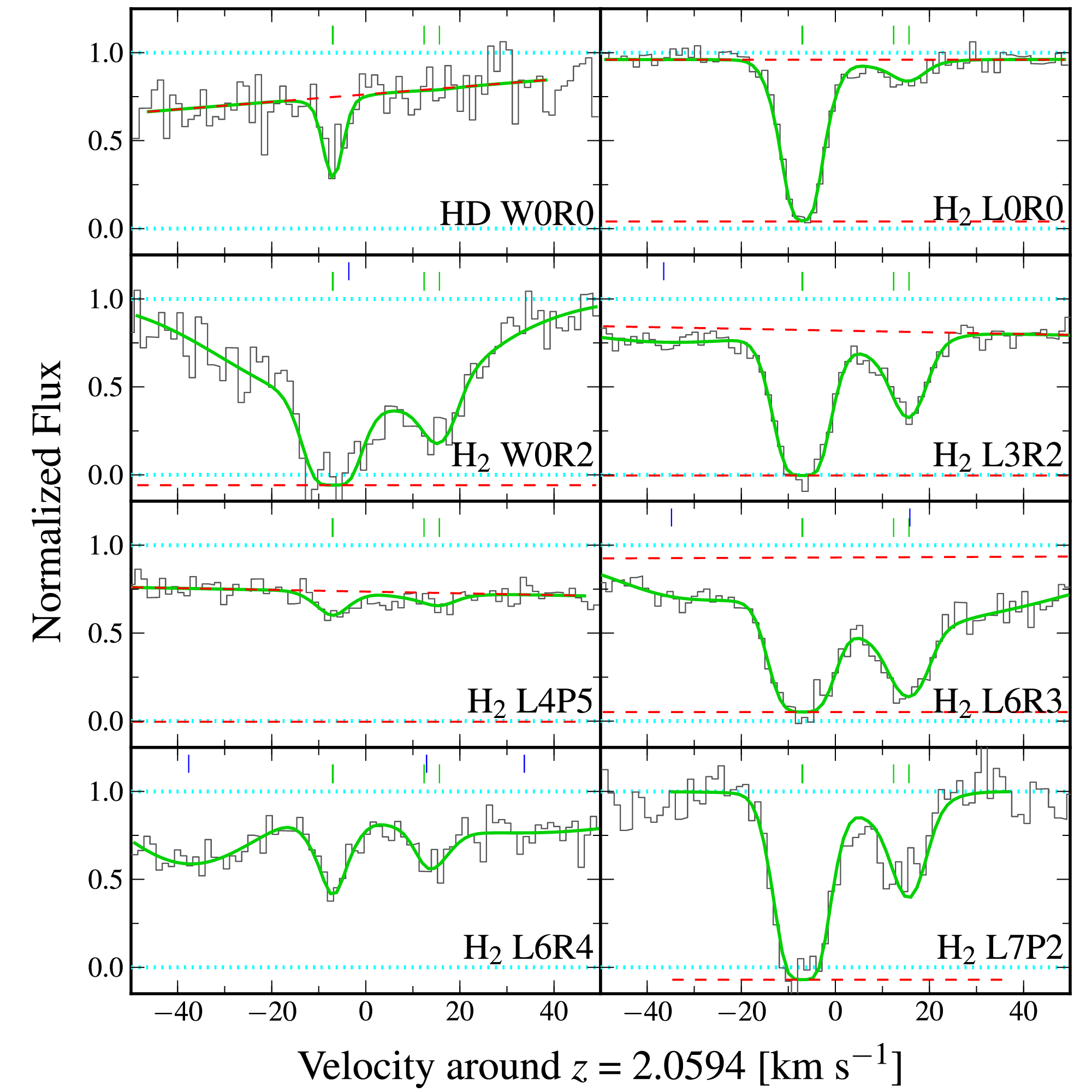}\vspace{-0.5em}
\caption{Some of the 86 H$_2$ and 7 HD lines from the J2123$-$0050
  Keck spectrum on a velocity scale centred at $z_{\rm abs}=2.0594$.
  The spectrum (black histogram) is normalized by a nominal continuum
  (upper dotted line). Local linear continua (upper dashed lines) and
  zero levels (lower dashed lines) are fitted simultaneously with the
  molecular and broader Lyman-$\alpha$ lines, the positions of which
  are indicated with, respectively, lighter and darker tick marks
  (offset vertically from each other) above the spectrum. The
  4-component fiducial fit is the solid curve. Note that the two
  left-most components are nearly coincident in velocity.}
\label{fig:lines}
\end{center}
\end{figure}

Figure \ref{fig:lines} highlights some of the H$_2$ and HD transitions
covering the observed range of overall line-strengths, {\SNR}s and
ground-state rotational levels, characterised by the quantum number
$J$ (`$J$-levels'). H$_2$ lines are observed in the
B$^1\Sigma_\mathrm{u}^+$--X$^1\Sigma_\mathrm{g}^+$ Lyman and
C$^1\Pi_\mathrm{u}$--X$^1\Sigma_\mathrm{g}^+$ Werner bands, for
$J\in[0,5]$. HD is observed in six R0 Lyman lines and one R0 Werner
line. The molecular absorption shows two distinct spectral features
(SFs), separated by $\approx20$\,\kms. Figure \ref{fig:fit_all} shows
that the strong left-hand SF appears saturated for most low-$J$
transitions, while the weaker right-hand SF appears unsaturated in
almost all transitions. Only the left-hand SF is detected in the HD
transitions. Clearly, accurately recovering $\Delta\mu/\mu$ with the
maximum precision available from the spectrum must take into account
this non-trivial absorption profile, the strength of which varies from
transition to transition, while also allowing the varying {\SNR}s,
degrees of Lyman-$\alpha$ blending, and possible continuum-placement
errors to contribute to the final uncertainty on $\Delta\mu/\mu$. We
achieve this with a simple $\chi^2$ fitting technique combined with a
detailed model of the absorption profile in Section \ref{ssec:chi}.

\subsection{Laboratory wavelengths for H$_2$ and HD transitions}\label{ssec:lab}

Equation (\ref{eq:shifts}) states that velocity shifts between
different H$_2$/HD transitions can be related to a varying $\mu$. But
it assumes that the current laboratory wavelengths of those
transitions are known to high enough precision to be considered exact
compared to the precision available from the astronomical spectra. And
with wavelengths $\lambda_{\rm lab}\la1150$\,\AA, laboratory
measurements of these transitions are challenging. Nevertheless, the
laboratory measurements have improved over the years and now provide
accurate and precise laboratory wavelengths for all the H$_2$/HD
transitions detected in our spectrum of J2123$-$0050 and, more
generally, for those relevant in $\mu$-variation studies.

While the classical data of the Meudon group reached fractional
wavelength accuracies of $\delta\lambda/\lambda\sim$10$^{-6}$
\citep{AbgrallH_93c}, the first laser calibration study using a pulsed
dye laser system \citep{HinnenP_94a} reached similar or somewhat
improved accuracies of $6\times10^{-7}$. The implementation of Fourier
transform-limited laser pulses in the harmonic conversion processes
\citep{UbachsW_97a} allowed an order of magnitude improvement. Using
this technique, several subsequent studies provided a data set of
Lyman and Werner transition wavelengths with fractional accuracies of
$5\times10^{-8}$ \citep{PhilipJ_04a,UbachsW_04a}. Recently, even more
accurate values -- fractional accuracies of $5\times10^{-9}$ for most
Lyman transitions and $1$--$2\times10^{-8}$ for Werner transitions --
were obtained from combining two-photon excitation results with
Fourier-transform studies of emission lines between H$_2$ excited
states \citep{SalumbidesE_08a}.

In our analysis we use the most accurate wavelength from these studies
for each H$_2$ transition detected in our spectrum of J2123$-$0050.
Table \ref{tab:h2} compiles the most precise laboratory data into a
complete, up-to-date catalogue for all allowed Lyman and Werner H$_2$
transitions between the lowest 8 rotational levels in the ground and
excited states with wavelengths generally above the hydrogen Lyman
limit (more specifically, to the first 20 and 6 excited vibrational
levels for Lyman and Werner transitions, respectively). Most important
to $\mu$-variation studies like ours are the wavelengths and
sensitivity coefficients (see below), but we have also included the
other laboratory data required for fitting the H$_2$ absorption lines
seen in astronomical spectra, such as oscillator strengths. And while
we detect only a small subset of these H$_2$ transitions in our
spectrum of J2123$-$0050, we include a complete list in Table
\ref{tab:h2} as a useful reference catalogue for other astronomical
studies.

The HD transitions were recently measured via direct extreme UV laser
excitation to a relative accuracy of $\approx5\times10^{-8}$ by
\citet{HollensteinU_06a} and \citet{IvanovT_08a} and we use those
values here. Table \ref{tab:hd} provides the laboratory data for the
10 HD transitions falling in our spectrum, 7 of which we use in our
analysis.

\begin{table}
\begin{center}
  \caption{Catalogue of the most accurate and precise laboratory
    parameters for fitting H$_2$ absorptions lines. Represented are
    all allowed Lyman and Werner H$_2$ transitions between the lowest
    8 rotational levels in the ground and excited states with excited
    state vibrational quantum numbers up to 20 and 6 for Lyman and
    Werner transitions, respectively. The first column provides a
    short-hand notation for the transition: letters denote a Lyman (L)
    or Werner (W) line and the branch, where P, Q and R represent
    $J^\prime-J=-1$, 0 and 1, respectively, for $J$ and $J^\prime$ the
    ground state and excited state $J$-levels, respectively; the first
    integer is the excited state vibrational quantum number and the
    second is $J$. The second column gives the most precise reported
    laboratory wavelength and its 1-$\sigma$ uncertainty and the third
    column provides the reference: 1 = \citet{BaillyD_09a}, 2 =
    \citet{UbachsW_07a} (a suffix ``a'' refers to directly measured
    wavelengths while ``b'' refers to wavelengths calculated from
    directly measured lines via combination differences) and 3 =
    \citet{AbgrallH_93c} for the excited state energy levels with
    ground states derived directly from \citet{JenningsD_84a}. Note
    that wavelengths with reference 3 are much less precise than those
    from references 1 and 2. The fourth column gives the oscillator
    strengths which were calculated from the Einstein $A$ coefficients
    given by \citet{AbgrallH_94a}.  The fifth column gives the
    (natural) damping coefficients which were calculated from the
    total transition probabilities ($A_{\rm t}$) in
    \citet{AbgrallH_00a}.  The final column gives the sensitivity
    coefficients calculated in \citet{UbachsW_07a} which have
    estimated uncertainties of typically $<5\times10^{-4}$ (see text).
    Only a small excerpt from the full table is presented here; the full
    table is available in the electronic version of this paper.}
\vspace{-0.5em}
\label{tab:h2}
\begin{tabular}{lcccccccc}\hline
Trans- & $\lambda_{\rm lab}$ & Ref. & $f$         & $\Gamma$           & $K$       \\
ition  & [\AA]               &      & [10$^{-2}$] & [10$^9$\,s$^{-1}$] &           \\\hline
L0P1   & 1110.062558(3)      &    1 & 0.05739     & 1.87               & $-0.0097$ \\
L0P2   & 1112.495989(3)      &    1 & 0.06915     & 1.86               & $-0.0119$ \\
L0P3   & 1115.895530(3)      &    1 & 0.07381     & 1.86               & $-0.0149$ \\
L0P4   & 1120.248839(3)      &    1 & 0.07560     & 1.85               & $-0.0187$ \\
L0P5   & 1125.540690(5)      &    1 & 0.07577     & 1.84               & $-0.0233$ \\
L0P6   & 1131.753504(6)      &    1 & 0.07481     & 1.83               & $-0.0286$ \\
L0P7   & 1151.6384(14)       &    3 & 0.07299     & 1.82               & $-0.0345$ \\\hline
\end{tabular}
\end{center}
\end{table}

\begin{table}
\begin{center}
  \caption{Laboratory data for $J=0$ HD transitions falling in (but
    not necessarily detected or fitted in) our spectrum of
    J2123$-$0050. The columns have the same descriptions as in Table
    \ref{tab:h2} except for the following. The laboratory wavelength
    references are 4 = \citet{HollensteinU_06a} and 5 =
    \citet{IvanovT_08a}. The oscillator strengths ($f$) and damping
    coefficients ($\Gamma$) were calculated from the Einstein $A$
    coefficients and total transition probabilities ($A_{\rm t}$),
    respectively, given by \citet{AbgrallH_06a}. The sensitivity
    coefficients ($K$) were calculated by \citet{IvanovT_08a} and have
    estimated uncertainties of typically $<1.5\times10^{-4}$.}
\vspace{-1.5em}
\label{tab:hd}
\begin{tabular}{lcccccccc}\hline
Trans- & $\lambda_{\rm lab}$ & Ref. & $f$         & $\Gamma$           & $K$                 \\
ition  & [\AA]               &      & [10$^{-2}$] & [10$^9$\,s$^{-1}$] &                     \\\hline
L0R0   & 1105.840555(57)     &    4 &  0.07436    & 1.87               &           $-0.0065$ \\
L1R0   & 1092.001264(58)     &    4 &  0.29669    & 1.76               &           $-0.0004$ \\
L2R0   & 1078.831044(61)     &    4 &  0.67473    & 1.67               & $\phantom{+}0.0053$ \\
L3R0   & 1066.27568(6)       &    5 &  1.14500    & 1.58               & $\phantom{+}0.0106$ \\
L4R0   & 1054.29354(6)       &    5 &  1.63570    & 1.50               & $\phantom{+}0.0156$ \\
L5R0   & 1042.85005(6)       &    5 &  2.05477    & 1.43               & $\phantom{+}0.0201$ \\
L6R0   & 1031.91493(6)       &    5 &  2.35911    & 1.36               & $\phantom{+}0.0244$ \\
L7R0   & 1021.46045(6)       &    5 &  2.53491    & 1.30               & $\phantom{+}0.0283$ \\
L8R0   & 1011.46180(6)       &    5 &  2.61849    & 1.25               & $\phantom{+}0.0319$ \\
L9R0   & 1001.89413(6)       &    5 &  2.47212    & 1.19               & $\phantom{+}0.0353$ \\
W0R0   & 1007.29020(6)       &    5 &  3.25368    & 1.18               &           $-0.0039$ \\\hline
\end{tabular}
\end{center}
\end{table}

\subsection{Sensitivity coefficients, $K$}\label{ssec:K}

The $K$ coefficients used here were calculated via a semi-empirical
analysis \citep{UbachsW_07a}. In this approach a Dunham representation
of the ground and excited state level structure is derived from fits
to the accurately determined transition frequencies.  In the first
instance, the levels most heavily perturbed by rotational-electronic
interactions are excluded. Known mass-scaling laws for the Dunham
coefficients then yield the $K$ coefficients. The non-adiabatically
perturbed levels were treated in the second instance, taking into
account wave function mixing of levels in the B$^1\Sigma_\mathrm{u}^+$
and C$^1\Pi_\mathrm{u}$ systems. Small adiabatic corrections (at the 1
per cent level) were included as well.  We note here that the values
reported in \citet{UbachsW_07a} include additional mass-dependent
effects on the non-adiabatic interaction matrix elements that were not
included in the initial study of \citet{ReinholdE_06a}; differences
between the two are nevertheless marginal. We also note that these $K$
coefficients are in very good agreement with the independent \emph{ab
  initio} calculations by \citet{MeshkovV_06a}; the agreement within 1
per cent is reassuring for both methods, since it corresponds to the
estimated uncertainties in both. The $K$ coefficients for the H$_2$
transitions are included in Table \ref{tab:h2}. The $K$ coefficients
for HD were derived in \emph{ab initio} calculations by
\citet{IvanovT_08a} and are given in Table \ref{tab:hd}.

\begin{figure}
\begin{center}
\includegraphics[width=\columnwidth]{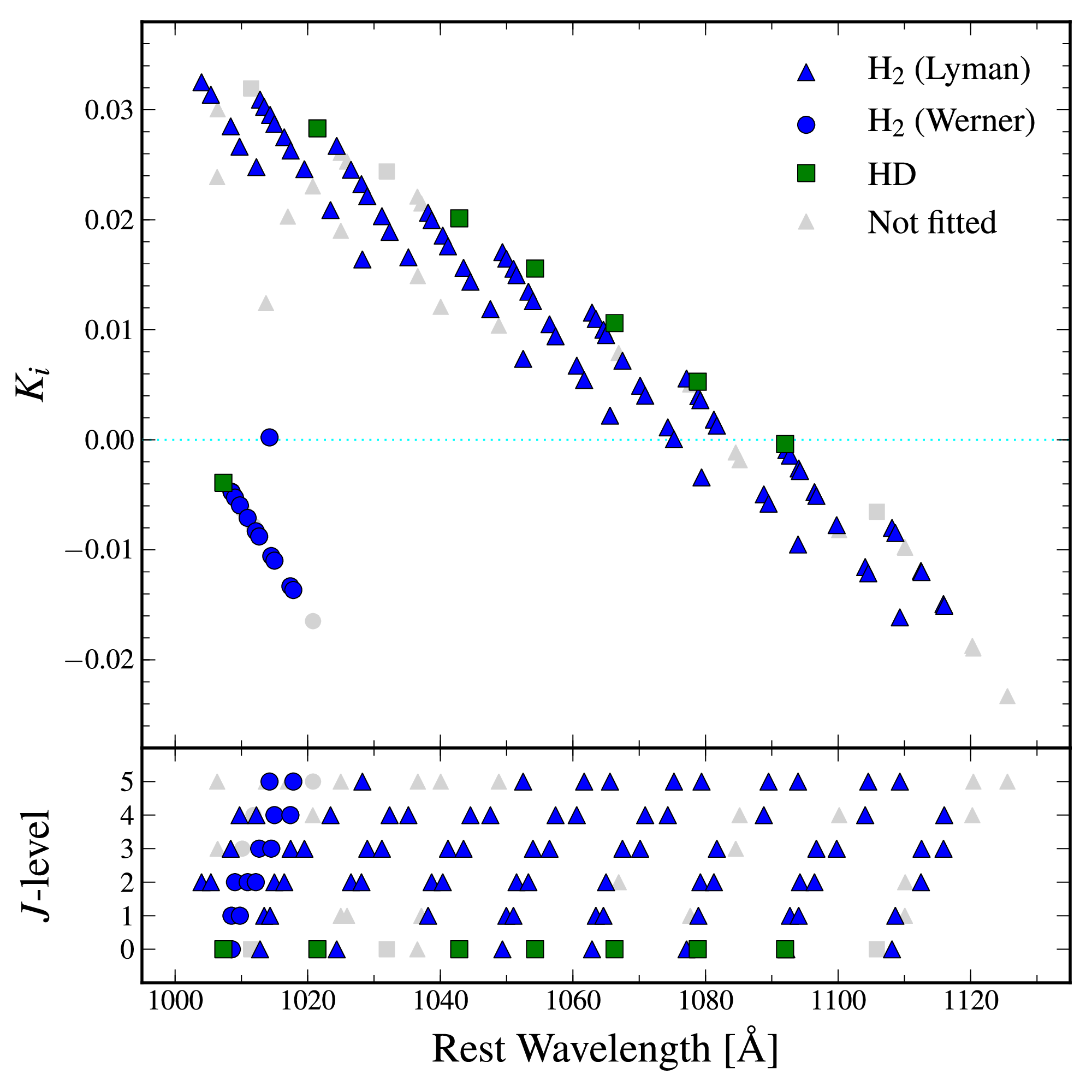}\vspace{-0.5em}
\caption{\emph{Upper panel:} The sensitivity coefficients, $K_i$, for
  the $J$=0--5 HD and H$_2$ Lyman and Werner transitions, $i$, used in
  our analysis (dark/coloured points) and those not detected or fitted
  (light grey points). The legend explains the symbols used in both
  panels. \emph{Lower panel:} The distribution of transitions with
  wavelength according to their $J$-levels.}
\label{fig:K}
\end{center}
\end{figure}

Figure \ref{fig:K} (upper panel) shows the $K$ values, which occupy
the range $-0.02<K<+0.03$, for all detected molecular transitions.
Note the general increase in $K$ for bluer Lyman transitions. Adding
the Werner transitions below $\lambda_{\rm lab}=1020$\,\AA\ provides a
more complicated signature of a varying $\mu$ because they shift in
the opposite direction to the Lyman transitions at similar
wavelengths.  These aspects of Fig.~\ref{fig:K} are particularly
important when considering possible systematic effects in Section
\ref{ssec:systematics}.

\section{Analysis and results}\label{sec:analysis}

\subsection{$\chi^2$ minimization analysis}\label{ssec:chi}

If we knew \emph{a priori} that molecular absorption occurred at a
single redshift in a single cloud and no other absorption lines were
present, measuring $\Delta\mu/\mu$ would be straightforward: each
molecular line's redshift could be determined from a fit against a
well-defined background continuum and its relative deviation from some
arbitrary redshift could be plotted against its $K$ coefficient.
Equation (\ref{eq:shifts}) implies that $\Delta\mu/\mu$ would simply
be the slope of such a plot. The offset would be degenerate with the
arbitrary redshift itself (though it clearly should be fitted as a
free parameter).

In reality, two main complicating factors are important. Firstly, all
molecular lines fall in the Lyman-$\alpha$ forest, the series of
comparatively broad H{\sc \,i} Lyman-$\alpha$ absorption lines
randomly distributed in optical depth and redshift bluewards of the
quasar's Lyman-$\alpha$ emission line. As Figs.~\ref{fig:fit_all} and
\ref{fig:lines} show, many molecular lines fall within the absorption
profile of one or more forest lines, so the latter must also be fitted
to provide an effective background. Secondly, the molecular lines show
`velocity structure' -- several absorbing clouds with similar
redshifts and different optical depths and Doppler widths. The two
main spectral features (SFs) highlighted in Fig.~\ref{fig:lines} are
obvious examples, but we demonstrate below that each SF comprises more
than one absorbing `velocity component' (VC).

The `simultaneous fitting' technique we employ here optimally and
robustly addresses these complications by fitting all detected H$_2$
and HD absorption profiles, together with the broader Lyman-$\alpha$
profiles, simultaneously in a single, comprehensive fit. Minimizing
$\chi^2$ allows uncertainties in the forest parameters -- the
effective continuum -- to contribute naturally to uncertainties in
$\Delta\mu/\mu$. Other advantages this technique brings over the
earlier `line-by-line' fitting approach are outlined by
\citet{KingJ_08a}.

\subsubsection{Free parameters and physical assumptions}\label{sssec:params}

Each absorption line is modelled as a Voigt profile -- the convolution
of Gaussian Doppler broadening and the transition's natural
line-shape. The oscillator strength and natural line-width determine
the latter; their values are provided in Tables \ref{tab:h2} and
\ref{tab:hd}. Three parameters describe each absorption cloud's
properties: Doppler width, $b$, column density, $N$, and redshift,
$z_{\rm abs}$. This is the only consideration for the Lyman-$\alpha$
lines but different molecular transitions share these parameters in
physically important ways.

For H$_2$ and HD, each $J$-level has a different ground state
population, so all transitions from the same $J$-level have the same,
single value of $N$ in each VC. This is clearly also true for the $b$
and $z_{\rm abs}$ values. However, we make the further assumption that
the velocity structure is the same in all $J$-levels, i.e.~that a
given VC has the same $z_{\rm abs}$ value in all $J$-levels. We test
the importance of this assumption in Section \ref{ssec:systematics}
and find that it has little impact on our results. We also assume that
a given VC is characterised by the same, single value of $b$ in all
$J$-levels; we test the importance of this assumption in Section
\ref{ssec:consistency}. In relating the absorption cloud parameters of
different transitions in these physically meaningful ways, the number
of free parameters is minimized. This is similar to the analysis in
\citet{KingJ_08a}, though they fitted the molecular oscillator
strengths as free parameters as well. Finally, because the HD
transitions are few and relatively weak -- recall that only the
left-hand spectral feature is detected -- we further assume that the
ratio of an HD VC's column density to that of the corresponding VC in
the H$_2$ $J$=0 transitions is the same for all VCs. As can be seen
from Fig.~\ref{fig:fit_all}, there is no strong evidence against this
assumption, a conclusion corroborated by removing the HD transitions
from the analysis altogether in Section \ref{ssec:consistency}.

There are several Fe{\sc \,ii} transitions with $\lambda_{\rm
  lab}<1150$\,\AA\ near some H$_2$ transitions which, in the $z_{\rm
  abs}=2.059$ absorber, appear quite weak but with velocity structure
extending over $\sim$400\,\kms. The velocity components in these
Fe{\sc \,ii} transitions are marked in Fig.~\ref{fig:fit_all}.  This
velocity structure was strongly constrained by using the Fe{\sc
  \,ii}\,$\lambda$1608\,\AA\ transition, which falls redwards of the
Lyman-$\alpha$ forest and so has a well-defined continuum. This
transition was blended with weak, broad C{\sc \,iv}\,$\lambda$1550
absorption associated with the quasar itself (Hamann et al., in
preparation). The C{\sc \,iv}\,$\lambda$1550 absorption was
constrained by fitting the corresponding C{\sc \,iv}\,$\lambda$1548
absorption. Thus, it was necessary to model both metal line species,
as shown in Fig.~\ref{fig:fit_all}, to obtain a reliable model of the
Fe{\sc \,ii} transitions appearing near our H$_2$ transitions of
interest. The H$_2$ fits are insensitive to those of the metal lines.

As noted in Section \ref{ssec:keck}, a nominal continuum was initially
fitted to the Lyman-$\alpha$ forest region before establishing a
detailed fit to the absorption lines. Simple, low-order polynomial
fits connecting seemingly unabsorbed regions of the spectrum,
typically spaced apart by $\ga$5000\,\kms, provided a qualitatively
realistic continuum. However, \emph{quantitatively} different continua
were determined by different authors of this paper; the differences
were large enough to slightly affect fits to individual absorption
lines. And because our line fitting approach explicitly connects the
properties of molecular lines to each other, it is important to
recognise that this nominal, \emph{fixed} quasar continuum is unlikely
to allow very good fits to the many molecular transitions
simultaneously. It is therefore important in such cases to fit local
continua around transitions where the nominal continuum is
particularly uncertain. These local continua are shown in
Figs.~\ref{fig:fit_all} and \ref{fig:lines}.

The reader will note that, for molecular transitions around which
broad Lyman-$\alpha$ lines are also fitted, there will be strong
degeneracies between the Lyman-$\alpha$ line parameters and those of
the local continuum. For this reason, the local continua have
polynomial degree of, at most, unity (i.e.~they are either constants
or straight lines in wavelength space). Also, the fitting code we use
indicates cases where the degeneracies are so strong that a local
continuum fit serves the same purpose as the Lyman-$\alpha$ line(s).
It should be emphasised that these ambiguities and uncertainties in
the local continuum shape all contribute to the uncertainties in the
molecular line fits in a natural way because, in our $\chi^2$
minimization process, all parameters of the fit are varied
simultaneously. Most previous works \citep[excluding,
e.g.,][]{KingJ_08a} effectively determined fixed, local continua
before fitting the molecular lines. This means that uncertainties in
the continuum fits do not propagate through to the more interesting
parameters of the subsequent fit, like $\Delta\mu/\mu$. While only a
small effect, the uncertainty in $\Delta\mu/\mu$ would be
underestimated with such an approach.

Weak night sky emission is nominally subtracted during the initial
flux extraction process, but systematic uncertainties in the zero flux
level can and do remain. There are many obvious examples in
Fig.~\ref{fig:fit_all} of saturated Lyman-$\alpha$ features which have
non-zero average flux (either positive or, usually, negative) in their
flat-bottomed line cores (e.g.~features near 3107, 3137, 3171, 3317
and 3350\,\AA). We therefore included the zero level as a free
parameter when fitting regions of spectrum which included nearly
saturated absorption lines (either Lyman-$\alpha$ and/or molecular).

We add to our list of free parameters the oscillator strengths for the
10 H$_2$ transitions with $J$=0--3 falling in the wavelength range
$\lambda_{\rm obs}=3345$--$3415$\,\AA. While establishing our fit,
these transitions had noticeably higher optical depths than predicted
by the model.  For this reason the oscillator strengths for these
transitions were left as free parameters to be determined in the
$\chi^2$ minimization process. Their fitted-to-calculated oscillator
strength ratios, $f_{\rm fit}/f_{\rm calc}$, determined using our
fiducial model (see below), range from 1.5 to 2.0. Although it is
possible that the published oscillator strengths for these
transitions, which (along with all other molecular lines used in our
fit) are calculated rather than experimentally measured, are
incorrect, we consider this very unlikely because these transitions
are not affected by interference between multiple interacting H$_2$
states. Fits to H$_2$ transitions, which included some of these
particular transitions, in other quasar absorbers also suggests their
published oscillator strengths are reliable (J.~King, private
communication). An alternative explanation is motivated by the fact
that these transitions all fall on the combined O{\sc
  \,vi}/Lyman-$\beta$ quasar emission line at $\lambda_{\rm
  lab}\approx1025$--$1045$\,\AA: perhaps there is some inhomogeneity
in the H$_2$ column density on the scale of the quasar broad
emission-line region. Independent evidence for this scenario does not
appear readily available in our spectrum, so it is difficult to
confirm and will be explored in future work. Nevertheless, given this
possible explanation, it is important to test whether these
transitions affect our final results, and we conduct this consistency
check in Section \ref{ssec:consistency}.

After establishing the best-fitting values of the model parameters
described above, we introduce the last free parameter -- the one of
primary interest here -- $\Delta\mu/\mu$. Note that only a single
value of $\Delta\mu/\mu$ describes all molecular transitions,
including all their constituent VCs. Equation (\ref{eq:shifts}) states
that a change in $\mu$ manifests itself as a pattern of relative
shifts between the different molecular transitions. Note that
Fig.~\ref{fig:K} shows that the diversity of $K$ values ensures
little covariance between $\Delta\mu/\mu$ and other parameters; the
pattern of shifts is not degenerate with, for example, the individual
VC redshifts.

\subsubsection{Minimizing $\chi^2$}\label{sssec:chi2}

The absorption model is constructed with the free parameters described
above and the $\chi^2$ between it and the spectral data is minimized
using the program {\sc
  vpfit}\footnote{http://www.ast.cam.ac.uk/$\sim$rfc/vpfit.html.}.
This is a non-linear least-squares $\chi^2$ minimization package
written specifically for modelling complex, possibly interrelated
absorption lines with a series of Voigt profiles. The model fit is
convolved with an instrumental resolution function for comparison with
the real spectrum; we used a Gaussian function with ${\rm
  FWHM}=2.7$\,\kms\ to match the resolution inferred from the
extracted ThAr exposures. The software `understands' that many
molecular transitions can arise from the same $J$-level (i.e.~that
they have the same $N$) and the ability to link physically related
parameters, such as $b$ and $z_{\rm abs}$ values, across many
transitions is inherent in its design.

$\Delta\mu/\mu$ was recently added to the {\sc vpfit} code as a free
parameter. The operation of a varying $\mu$ on the molecular lines is
functionally the same as the action of a varying $\alpha$ on metal
lines. That {\sc vpfit} returns the correct values of
$\Delta\alpha/\alpha$ and its uncertainty has been tested with a
variety of simulations, including models with many strongly
overlapping VCs which are, in some cases, blended with transitions
from unrelated absorption clouds \citep{MurphyM_03a}. However, the fit
we conduct here is substantially more complicated (though
operationally similar) to such metal-line fits because the number of
transitions and links between fitted parameters is much larger.
Appendix \ref{app:a} describes a Monte Carlo test of {\sc vpfit} using
our actual fit to the molecular, Lyman-$\alpha$ forest and metal
absorption lines in J2123$-$0050 as a synthetic spectrum, to which
noise is added and into which a value of
$\Delta\mu/\mu=+5\times10^{-6}$ is inserted. The $\chi^2$ minimization
of the fit is then re-run on 420 such realizations. The mean and
distribution of the $\Delta\mu/\mu$ values recovered agrees well with
expectations. We also conducted convergence tests using our fit to the
real data to ensure that {\sc vpfit} arrived at the same value of
$\Delta\mu/\mu$ given a range of starting values; the default for all
the results quoted in this paper was zero, but starting values of,
e.g., $\pm10$ and $\pm5\times10^{-6}$ gave the same results. The
self-consistency checks conducted in Section \ref{sec:sys} also
corroborate the reliability of the algorithm.

The 1-$\sigma$ uncertainties on the best fitting parameters are
derived by {\sc vpfit} from the appropriate diagonal terms of the
final parameter covariance matrix. Given a particular absorption
model, these errors represent only the formal statistical
uncertainties derived from the flux error arrays of the fitted
spectral regions. They do not reflect additional sources of error,
like wavelength calibration uncertainties (see Section \ref{sec:sys}).
Nor do they represent `model errors' -- the possible systematic errors
involved with selecting an appropriate absorption model. We discuss
the selection of the `best' model at length in Section
\ref{ssec:fiducial}. {\sc vpfit} assumes that the fluxes (and flux
errors) in neighbouring spectral pixels are uncorrelated. This is
certainly not true in practice because the different quasar exposures
were re-binned onto the same final wavelength grid for combination
into a single spectrum. The correlation is reduced by averaging these
exposures. And we are unable to detect the effect by comparing the RMS
flux variations in unabsorbed regions with the final error spectrum;
these values are very similar. We demonstrate in Section
\ref{ssec:systematics} that the binning of the different exposures
onto the same wavelength grid produces a small systematic effect on
$\Delta\mu/\mu$.

\subsection{Fiducial absorption model}\label{ssec:fiducial}

Nearly all the H$_2$ and HD transitions detected in our spectrum of
J2123$-$0050 were utilized in our fit; some transitions' profiles were
heavily blended with either nearby Lyman-$\alpha$ lines and/or
suspected metal lines from one or more absorption systems (e.g.~the
L0P1 H$_2$ transition falling at $\lambda_{\rm obs}\approx3396.2$, see
Fig.~\ref{fig:fit_all}). No reliable fit could be obtained so we chose
not to include these molecular lines in our sample.

In total, 86 H$_2$ transitions and 7 HD transitions are
included in the fiducial fit. Only small regions around the molecular
lines were fitted: each region contained only enough of the
surrounding Lyman-$\alpha$ forest absorption to define the effective
continuum against which the molecular lines were fitted. This meant
that some molecular lines were fitted together in the same spectral
region, joined by fitted Lyman-$\alpha$ forest lines and, in most
cases, with a single (straight-line) local continuum and/or zero
level.  All the fitted regions are shown in Fig.~\ref{fig:fit_all}.
Assuming a simple, two VC model for the molecular absorption, an
initial model of the Lyman-$\alpha$ forest lines, local continua and
zero levels was established and refined.  This same basic structure
was used as a starting point in subsequent fits to determine the
molecular velocity structure.

\begin{figure}
\begin{center}
\includegraphics[width=\columnwidth]{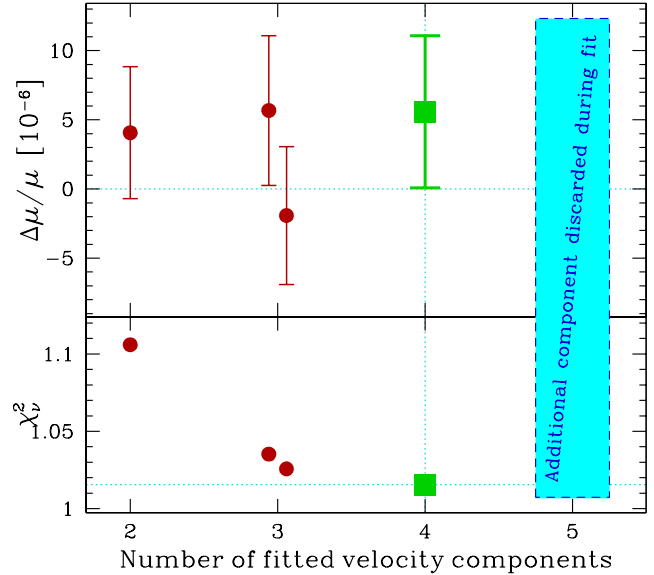}\vspace{-0.5em}
\caption{$\Delta\mu/\mu$ (upper panel) and $\chi^2$ per degree of
  freedom, $\chi^2_\nu$ (lower panel), for different velocity
  structures characterised by the number of fitted absorption
  components. The 4-component fit, highlighted with square points, has
  the lowest $\chi^2_\nu$ and is therefore the statistically preferred
  model. Two qualitatively different 3-component fits were possible;
  their results are horizontally offset here for clarity. Note the
  very different values of $\Delta\mu/\mu$ they return, exemplifying
  the inherent systematic effects associated with `under-fitting' the
  absorption profile. The error bars represent 1-$\sigma$ statistical
  uncertainties only; systematic errors are discussed in Section
  \ref{ssec:systematics}.}
\label{fig:comps}
\end{center}
\end{figure}

Although the absorption profiles in Fig.~\ref{fig:lines} clearly show
two spectral features, it is possible, even likely, that the profiles
comprise more than two VCs, each of which would be detectable (in the
absence of the others) given the large number of molecular transitions
we observe. The velocity structure best representing the molecular
absorption profile was determined by fitting models with increasing
numbers of molecular VCs and selecting the one with the smallest
$\chi^2$ per degree of freedom, $\chi^2_\nu$. This is one simple,
standard method for discriminating between different models in
$\chi^2$ analyses, though other (very similar) `information criteria'
can also be used \citep[e.g.][]{LiddleA_07a}. Note that while (the
minimum) $\chi^2$ itself must always decrease when more free
parameters are added to a model, $\chi^2_\nu$ begins to
\emph{increase} when the additional parameters are not statistically
justified. Figure \ref{fig:comps} (lower panel) shows $\chi^2_\nu$
versus the number of molecular VCs in the fit. The 4-component model
has a lower $\chi^2_\nu$ than the 3-component models and so is
statistically preferred. 5- and 6-component fits were attempted but
{\sc vpfit} found the additional components to be statistically
unnecessary. That is, during the $\chi^2$ minimization process those
components became so weak that $\chi^2$ became insensitive to their
parameters, so the components were removed. Equivalently, if they had
remained in the fit, the final, reduced $\chi^2_\nu$ would have been
larger than for the fiducial 4-component fit.

Figures \ref{fig:fit_all} and \ref{fig:lines} show our fiducial
4-component fit. Table \ref{tab:params} provides the molecular cloud
properties -- $z_{\rm abs}$, $b$ and the column densities in H$_2$ and
HD -- and their formal 1-$\sigma$ statistical uncertainties in all
relevant $J$-levels for each of the 4 VCs. Milutinovic et al.~(in
preparation) and Tumlinson et al.~(in preparation) consider the
absorption cloud properties in detail.

\begin{table*}
\begin{center}
  \caption{Molecular absorption line parameters and 1-$\sigma$
    statistical uncertainties for the 4-component fiducial fit in
    Figs.~\ref{fig:fit_all} and \ref{fig:lines}. Note the assumptions
    discussed in Section \ref{sssec:params} which minimise the number
    of free parameters. In particular, the ratio of the HD ($J=0$)
    velocity component column densities to those of the corresponding
    H$_2$ $J$=0 components is the same for all components. We also
    provide the total column density for each H$_2$ $J$ level and for
    HD.}
\label{tab:params}\vspace{-1.0em}
{\footnotesize\begin{tabular}{cccccccccc}\hline
Com-      & $z_{\rm abs}$  & $b$            & \multicolumn{6}{c}{$\log$$N($H$_2)$ [cm$^{-2}$]}                                                    & \multicolumn{1}{c}{$\log$$N($HD$)$}\\
ponent    &                & [km\,s$^{-1}$] & $J=0$          & $J=1$          & $J=2$          & $J=3$          & $J=4$          & $J=5$          & [cm$^{-2}$]                        \\\hline
1         & 2.0593276(5)   & $5.14\pm0.11$  & $15.06\pm0.03$ & $15.38\pm0.06$ & $15.01\pm0.03$ & $15.03\pm0.02$ & $13.98\pm0.04$ & $13.76\pm0.06$ & $12.95\pm0.03$                     \\
2         & 2.0593290(4)   & $1.92\pm0.06$  & $15.80\pm0.40$ & $17.52\pm0.04$ & $16.23\pm0.16$ & $15.16\pm0.25$ & $13.45\pm0.09$ & $12.78\pm0.37$ & $13.69\pm0.05$                     \\
3         & 2.0595264(76)  & $9.66\pm0.90$  & $13.90\pm0.08$ & $14.30\pm0.07$ & $14.10\pm0.05$ & $14.11\pm0.07$ & $13.31\pm0.22$ & $13.47\pm0.15$ &                                    \\
4         & 2.0595597(8)   & $4.02\pm0.16$  & $13.81\pm0.06$ & $14.69\pm0.03$ & $14.26\pm0.04$ & $14.39\pm0.04$ & $13.41\pm0.09$ & $13.22\pm0.15$ &                                    \\
Total     &                &                & $15.88\pm0.14$ & $17.53\pm0.04$ & $16.26\pm0.11$ & $15.46\pm0.04$ & $14.23\pm0.03$ & $14.04\pm0.04$ & $13.77\pm0.03$                     \\\hline
\end{tabular}}
\end{center}
\end{table*}

When visually assessing the fit in Figs.~\ref{fig:fit_all} and
\ref{fig:lines}, one clearly cannot readily ``see'' 4 VCs and so one
may worry that too many components are being fitted. However, this is
merely a failing of the human eye to recognise coherent structures in
almost 100 transitions simultaneously. For example, it is simple to
simulate a similar molecular absorption profile as in the real data
which comprises 4 components, one or more of which might be subtle
enough to go unnoticed when visually assessing several transitions but
which is statistically significant when $\sim$100 transitions are
analysed properly. And so we must appeal to the objective, well-tested
and well-understood model selection technique discussed above to {\it
  recover all the statistically significant structure in the line
  profiles}, otherwise we should expect systematic errors in our
results.  Previous works have demonstrated that `under-fitting' --
fitting too few VCs -- causes strong systematic errors in quantities
like $\Delta\alpha/\alpha$ \citep*{MurphyM_08a} and $\Delta\mu/\mu$
\citep{MurphyM_08b} when derived from individual absorption systems.

\subsection{Fiducial result}\label{ssec:result}

Given an absorption model for which $\chi^2$ has been minimized,
determining $\Delta\mu/\mu$ is straightforward: it is simply among the
many free parameters determined in the $\chi^2$ minimization process.
For the 4-component fiducial model, we find that
\begin{equation}\label{eq:raw_result}
\Delta\mu/\mu=(+5.6\pm5.5_{\rm stat})\times10^{-6}\,.
\end{equation}
Again, we emphasise that the 1-$\sigma$ uncertainty derives only from
the spectrum's photon statistics and is calculated from the relevant
diagonal term of the final parameter covariance matrix. Given the
range of $K$ coefficients involved ($\sim$0.05; see
Fig.~\ref{fig:K}), equation (\ref{eq:shifts}) implies that this
uncertainty corresponds to a velocity precision of $\sim$80\,\ms, or
$\sim$0.06 spectral pixels.

Figure \ref{fig:comps} (upper panel) shows the values of
$\Delta\mu/\mu$ for different models of the molecular velocity
structure as characterised by the number of fitted absorption
components. It graphically illustrates the dangers of `under-fitting'.
The two different 3-component fits possible in this system give quite
different $\Delta\mu/\mu$ values. This large difference should be
placed into the context that, since all fits were carried out on the
same transitions in the same spectrum, the results from the different
models are by no means independent; the statistical error bars are not
the gauge of ``significant'' differences here. Also, the 2-component
model may seem statistically acceptable at first since its
$\chi^2_\nu\sim1$. However, its $\chi^2_\nu$ is much larger than for
the 4-component fit and so it cannot be preferred in any objective
sense. This is also true of the two different 3-component fits.

\begin{figure}
\begin{center}
\includegraphics[width=\columnwidth]{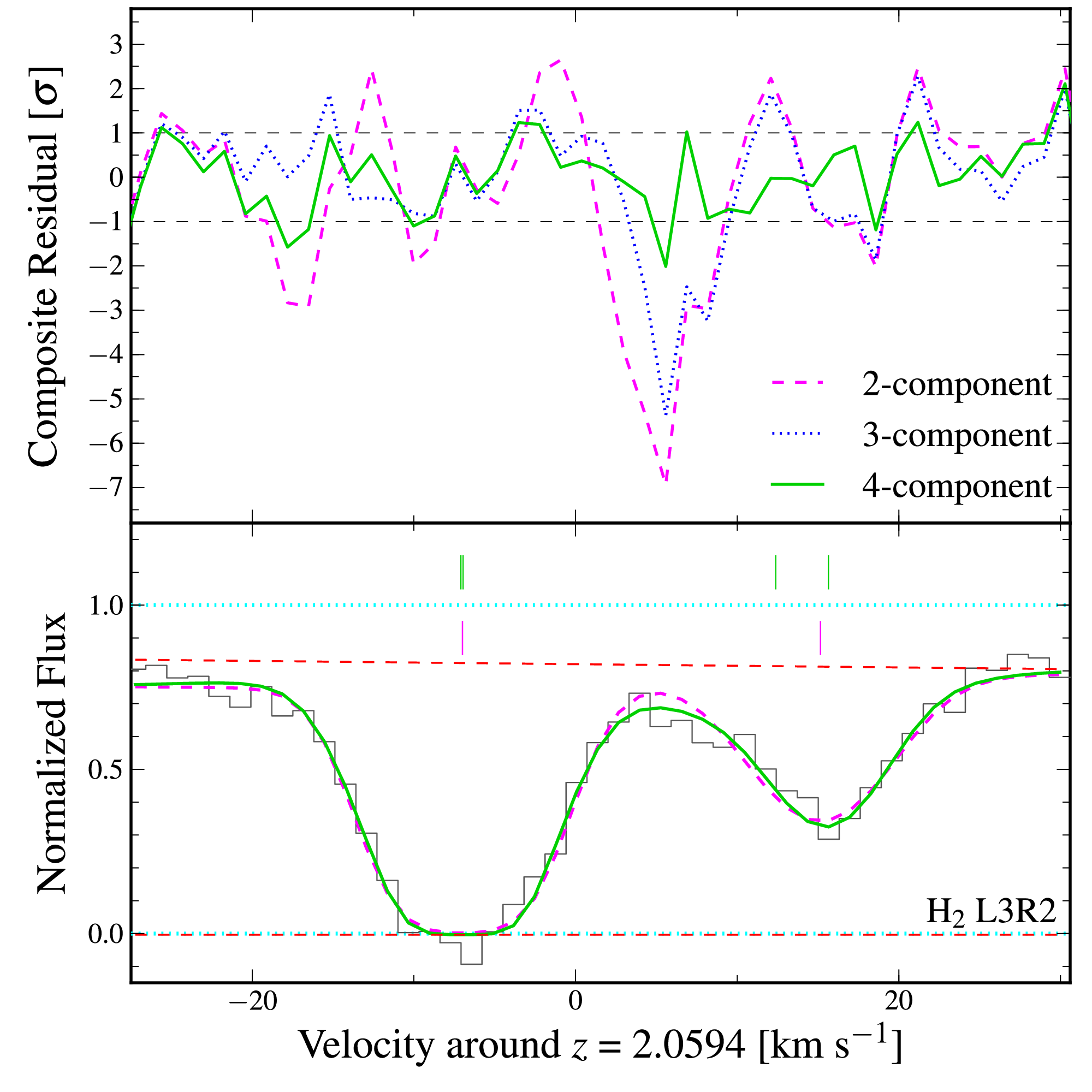}\vspace{-0.5em}
\caption{Composite residual spectra (CRS), formed from 24 relatively
  unblended H$_2$ transitions. \emph{Bottom panel}: An example H$_2$
  transition fitted with the 2- and fiducial 4-component models
  (dashed and solid curves, respectively).  Only a small difference is
  noticeable by eye. However, the CRS panel above shows how significant
  the additional components are.  \emph{Upper panel}: CRS for the
  4-component fiducial model (solid line) and the 2- and 3-component
  models (dashed and dotted lines, respectively; for clarity only one
  of the two possible 3-component models is shown here). Note the
  large, many-pixel excursions outside the $\pm$1-$\sigma$ range for
  the 2- and 3-component models, indicating unmodelled velocity
  structure. No such features exist for the 4-component model, as
  expected.}
\label{fig:crs}
\end{center}
\end{figure}

The inadequacies of 2- and 3-component fits are further exposed in
Fig.~\ref{fig:crs}. For the 24 relatively unblended H$_2$ transitions,
the residuals between the data and the model fit were normalized by
the flux error arrays, shifted to a common velocity scale and averaged
together. This forms a `composite residual spectrum' as a diagnostic
for under-fitting. By combining the residual structure between the
model and the data for many transitions, we may now better appreciate
``by eye'' how well the model describes the statistically significant
structure in those absorption profiles.  Figure \ref{fig:crs} clearly
shows that the 2-component model fails to reproduce the real
absorption profile shape, leaving many-pixel excursions outside the
expected residual range as evidence of additional VCs. This is also
true for the 3-component models. By comparison, the 4-component model
leaves no obvious evidence for unmodelled, statistically significant
structure.

\section{Internal consistency and systematic errors}\label{sec:sys}

\subsection{Consistency tests}\label{ssec:consistency}

The fiducial model in Section \ref{ssec:fiducial} includes all
detected molecular transitions which could be fitted reliably; the
data quality and degree of Lyman-$\alpha$ blending naturally weights
each one's contribution to $\Delta\mu/\mu$. The main advantage of this
approach is objectivity: we do not select which transitions are `best'
to fit, or those which we expect, \emph{a priori}, to provide the
strongest constraints on $\Delta\mu/\mu$. Another approach is to fit
only the `least blended' or the `well-defined' transitions. Deciding
which transitions fit these descriptions is clearly subjective.
Nevertheless, as a consistency check, we fitted spectral regions
containing either no detectable Lyman-$\alpha$ blending or very simple
and weak blending. We also excluded transitions blended with the
Fe{\sc \,ii} lines described in Section \ref{sssec:params}. With 53
H$_2$ and 5 HD remaining transitions, a 4-component model was
statistically preferred, returning
$\Delta\mu/\mu=(+5.0\pm5.9)\times10^{-6}$.

The $J=4$ and $5$ H$_2$ transitions are weak, typically absorbing
$<30$\% of the local continuum. It is possible, though we would argue
unlikely, that unmodelled structure in the nearby Lyman-$\alpha$
forest lines and/or local continua for these transitions will have a
greater relative effect on the molecular profile fits in these cases.
A 4-component fit where the 28 $J=4$ and $5$ transitions are prevented
from directly influencing $\Delta\mu/\mu$, effectively by setting
their $K$ values to zero\footnote{At first it may seem that a
  simpler, more easily interpretable alternative would be to
  completely remove the $J$=4 and 5 (or HD) transitions to test their
  influence on the fiducial result. However, as we describe in Section
  \ref{sssec:params} and as is apparent in Fig.~\ref{fig:fit_all},
  several molecular transitions are often fitted together in one
  contiguous fitting region. This is because different transitions can
  be blended together, or be effectively joined by the same
  Lyman-$\alpha$ forest lines. Thus, in principle, removing individual
  molecular transitions from the fit is not possible without removing
  entire fitting regions.}, yielded
$\Delta\mu/\mu=(+4.8\pm5.6)\times10^{-6}$. The very small increase in
the statistical uncertainty compared to equation (\ref{eq:raw_result})
is symptomatic of how weakly these $J=4$ and $5$ transitions affect
$\Delta\mu/\mu$, due primarily to their low optical depth. A similar
problem may exist for the HD transitions, which are also very weak; a
4-component model with the HD transitions' dependence on $\mu$ removed
returned $\Delta\mu/\mu=(+4.1\pm5.8)\times10^{-6}$.

In Section \ref{sssec:params} we described how the predicted optical
depths for the $J$=0--3 H$_2$ transitions falling at $\lambda_{\rm
  obs}=3345$--$3415$\,\AA\ (i.e.~where the Lyman-$\beta$/O{\sc \,iv}
quasar emission line falls) were too low, possibly because there is
some inhomogeneity in the H$_2$ column density on the angular scale of
the quasar broad line region. We therefore allowed these transitions'
oscillator strengths to effectively vary as free parameters in the
fit. This may not be an accurate parametrisation so we must assess the
extent to which it affects our conclusions. Firstly, if we remove the
$\mu$-dependence of the 10 $J$=0--3 H$_2$ transitions concerned (by
setting their $K$ values to zero), a 4-component fit gives
$\Delta\mu/\mu=(+7.0\pm6.7)\times10^{-6}$. Another test is to fit all
H$_2$/HD transitions with oscillator strengths as free parameters,
i.e.~the column densities of different transitions from the same
$J$-level are allowed to be different, as in the fits conducted by
\citet{KingJ_08a}. A 4-component fit yields
$\Delta\mu/\mu=(+5.4\pm5.5)\times10^{-6}$. Notice that the statistical
uncertainty does not increase discernibly here, even though the number
of fitted parameters has been increased significantly; this is because
the many column-density parameters for the molecular velocity
components of each transition have very little influence on the fitted
line positions, which is what primarily affects $\Delta\mu/\mu$.

Our fiducial model assumes that a given molecular VC has the same
$b$-parameter in all $J$-levels. One can envisage different physical
effects which may invalidate this model assumption. And while we do
not find obvious evidence to the contrary in our spectra, it may be
that our fitting approach -- where we make the assumption and then
build up the fit to the molecular lines by adding VCs until all the
statistical structure in the molecular line profiles is modelled -- is
not the best method for identifying such an effect. We have therefore
relaxed the assumption to test its effect on the fitted value of
$\Delta\mu/\mu$: when allowing VCs to have different $b$-parameters in
different $J$-levels, $\Delta\mu/\mu=(+7.9\pm5.6)\times10^{-6}$. While
there is small change in the value of $\Delta\mu/\mu$ with respect to
the fiducial value in equation (\ref{eq:raw_result}), it does not
represent a strong model dependency. We note in passing that allowing
VCs to have different $b$-parameters in different $J$-levels does not
solve the problem, discussed above, of some H$_2$ transitions appearing
stronger in the spectrum compared to the prediction of the fiducial
fit; those transitions come from a range of different $J$-levels.

\begin{figure}
\begin{center}
\includegraphics[width=\columnwidth]{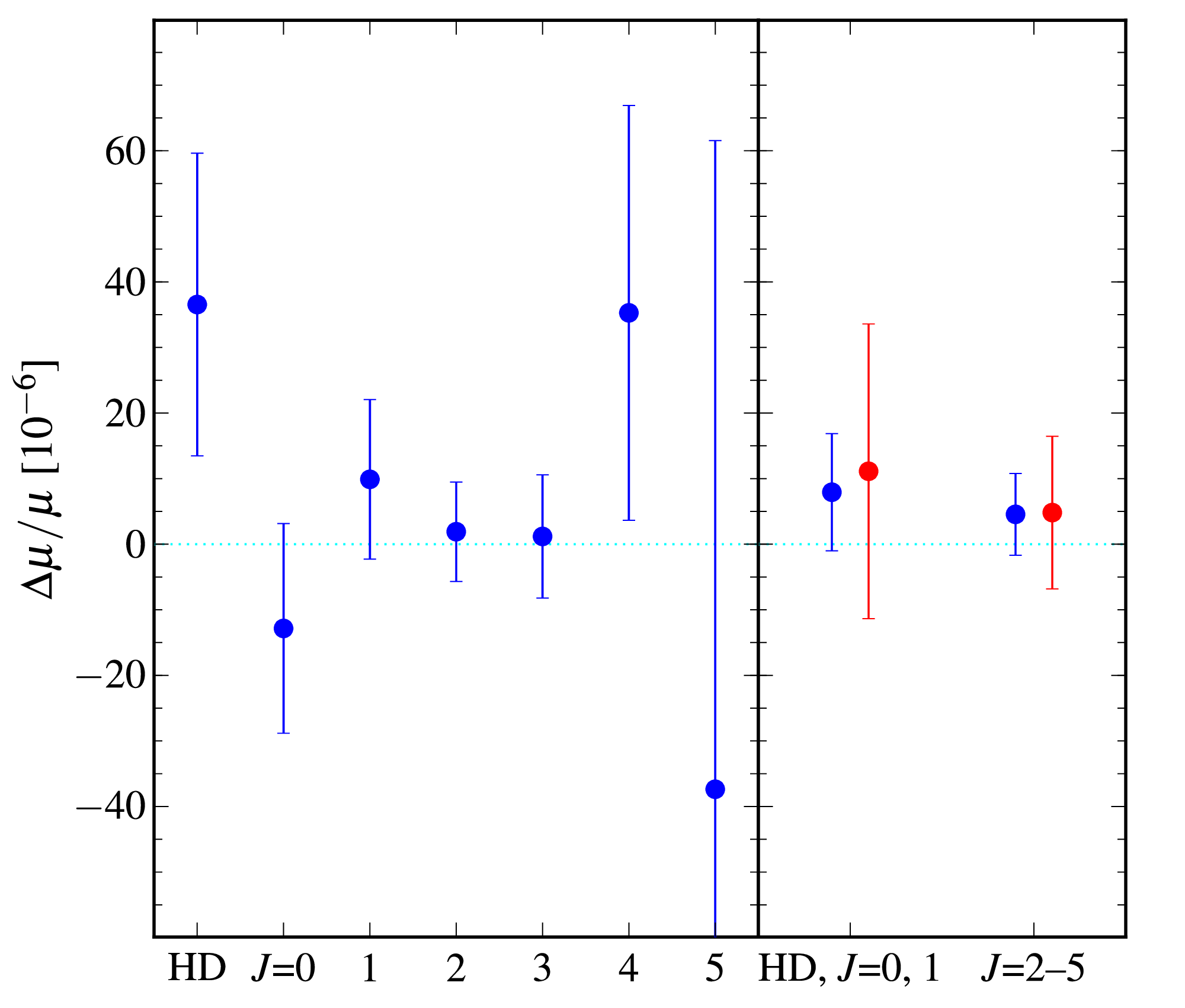}\vspace{-0.5em}
\caption{\emph{Left panel}: $\Delta\mu/\mu$ for each H$_2$ $J$-level
  and HD, assuming they all have the same 4-component velocity
  structure. \emph{Right panel}: $\Delta\mu/\mu$ for two groups of
  transitions, HD plus H$_2$ $J$=0--1 and H$_2$ $J$=2--5. The
  dark/blue points derive from a model assuming the same velocity
  structure in both groups, whereas the light/red points (offset to
  the right for clarity) allow the velocity structure for the two
  groups to differ.  Both groups were fitted with a 4-component model
  in both cases.}
\label{fig:J}
\end{center}
\end{figure}

Finally, {\sc vpfit} allows different values of $\Delta\mu/\mu$ to be
fitted for the different H$_2$ $J$-levels. In this test we maintain
the assumption that all $J$-levels have the same velocity structure,
i.e.~all $J$-levels are fitted with 4 VCs, with each VC having the
same redshift in all transitions (we relax this assumption in Section
\ref{ssec:systematics}). We also derived a separate value from the HD
lines. The resulting values of $\Delta\mu/\mu$ are plotted in
Fig.~\ref{fig:J} (left panel). This plot is useful in clarifying the
relative contributions the series of transitions from different H$_2$
$J$-levels and HD make to the final uncertainty on $\Delta\mu/\mu$.
Of course, the velocity structure was assumed to be the same in all
cases, so the different $\Delta\mu/\mu$ values are not strictly
independent. Nevertheless, we note general agreement, especially for
the H$_2$ $J$=1--3 transitions. Since the error bars are relatively
large for the other $J$-levels and HD, we also derived less uncertain
values by grouping the HD and H$_2$ $J$=0--1 transitions together.  A
similar grouping of the H$_2$ $J$=2--5 transitions gave a similar
value of $\Delta\mu/\mu$. These values are plotted on Fig.~\ref{fig:J}
(right panel).

The above consistency checks indicate that the fiducial value of
$\Delta\mu/\mu$ is not strongly dependent on the assumptions made in
our 4-component fit. Individually relaxing these assumptions produces
little change in the measured value of $\Delta\mu/\mu$. This is also
consistent with there being no important problems with our data, or
our analysis.

\subsection{Systematic errors}\label{ssec:systematics}

\subsubsection{Specific, known wavelength calibration
  errors}\label{sssec:wcal_specific}

Since a varying $\mu$ should cause velocity shifts between molecular
transitions, wavelength calibration errors which \emph{distort} the
wavelength scale are clearly important potential sources of systematic
errors in $\Delta\mu/\mu$. Note that, as implied by equation
(\ref{eq:shifts}), effects which shift the wavelength scale by a
constant velocity at all wavelengths are generally
unimportant\footnote{This is not strictly true when many quasar
  exposures are combined, as is the case here. If different velocity
  shifts are applied to the different quasar exposures, and if the
  relative weights of the exposures vary with wavelength when forming
  the final, combined spectrum, then small relative velocity shifts
  will be measured between transitions at different wavelengths.}. The
first, most obvious possibility is that the wavelength scale of the
ThAr exposures was incorrectly determined. This is easily checked. The
wavelength calibration residuals, with ${\rm RMS}\sim80$\,\ms, are
consistent with being symmetrically distributed around the final
wavelength solution at all wavelengths. Because many H$_2$/HD
transitions are used, with a range of wavelengths, this reduces the
overall calibration error substantially. For example, over the 12
HIRES echelle orders covering the wavelength range where the molecular
lines fall (3071--3421\,\AA), 150--200 ThAr lines are used to
calibrate our ThAr exposures. Clearly though, systematic trends in
these residuals are still possible. Using the same technique employed
in \citet{MurphyM_07b} to track systematic patterns in the ThAr
calibration residuals, the possible distortion in the wavelength scale
between 3070 and 3430\,\AA\ is $<$30\,m\,s$^{-1}$. This corresponds to
a systematic error in $\Delta\mu/\mu$ of $\pm2.0\times10^{-6}$ at
most.

Once determined from the ThAr exposures, the wavelength solutions are
simply applied to the corresponding quasar exposures. Drifts in the
refractive index of air inside HIRES between the ThAr and quasar
exposures will therefore cause miscalibrations. The temperature and
atmospheric pressure drifts during our observations were $<$1\,K and
$<1$\,mbar respectively. According to the \citet{EdlenB_66a} formula
for the refractive index of air, this would cause \emph{differential}
velocity shifts between 3070 and 3430\,\AA\ of $<$10\,m\,s$^{-1}$,
which are negligible.

The quasar and ThAr light traverse similar but not identical paths
through HIRES. Only the quasar light passes from the telescope into
HIRES and the ThAr light (nearly) uniformly illuminates the slit while
the quasar light is centrally concentrated. \emph{Different}
distortions of the wavelength scale -- within individual echelle
orders (`intra-order') and over longer wavelength ranges
(`long-range') -- may therefore occur in the quasar and corresponding
ThAr exposures.

\citet{GriestK_10a} recently identified intra-order distortions in
HIRES. They compared the wavelength scales established using ThAr
exposures with those imprinted on quasar exposures by an iodine
absorption cell. The distortions are such that, for different
transitions at the same redshift in a quasar spectrum, those at the
order edges appear at positive velocities with respect to transitions
at the order centres when calibrated with a ThAr exposure. The
peak-to-peak velocity distortion is $\sim$500\,\ms\ at
$\sim$5600\,\AA\ and may worsen (improve) with increasing (decreasing)
wavelength; see figure 4 of \citet{GriestK_10a}. The effect is similar
(though not identical) for all echelle orders in the wavelength range
$\sim$5000--6200\,\AA\ covered by the iodine cell absorption. If we
assume that similar intra-order distortions apply to our spectra of
J2123$-$0050 at much bluer wavelengths then, because the molecular
transitions of interest lie at different positions along different
echelle orders, we should expect the effect on $\Delta\mu/\mu$ to be
suppressed. To illustrate this, a very crude estimate of the
`residual' velocity distortion is just the observed $\sim$500\,\ms\
peak-to-peak intra-order value reduced according to the number of
molecular transitions observed, i.e.~$\sim500\,\ms/\sqrt{93} \approx
52\,\ms$. This corresponds to a systematic error in $\Delta\mu/\mu$
of approximately $\pm3.5\times10^{-6}$.

This crude calculation fails to take into account the strong variation
in the spectral \SNR\ across the wavelength range containing the
H$_2$/HD transitions, their different sensitivities to $\mu$
variation, where they fall with respect to echelle order edges in our
particular spectrum and other factors such as the degree of blending
with Lyman-$\alpha$ forest lines of each molecular transition. A
simple Monte Carlo simulation which takes into account the first two
of these effects -- \SNR\ and $K$ for each transition -- was
undertaken as follows. Each realisation comprised a velocity shift,
$\Delta v_i$, chosen randomly for each transition, $i$, from the
500\,\ms\ peak-to-peak interval spanned by the intra-order distortions
identified by \citet{GriestK_10a}. The square of the \SNR\ of the
spectral data surrounding the molecular transitions then weighted a
linear least squares fit of the $\Delta v_i/c$ versus $K$ values, the
slope of which provided a value of $\Delta\mu/\mu$ for that
realisation -- see equation (\ref{eq:shifts}). The distribution of
values for $\Delta\mu/\mu$ over hundreds of thousands of Monte Carlo
realizations was close to Gaussian with an RMS of $4.8\times10^{-6}$.

However, the most direct calculation of the possible systematic effect
on $\Delta\mu/\mu$, which takes into account all remaining effects
(e.g.~positions of transitions with respect to order edges, line
blending etc.), can be obtained by attempting to remove the
intra-order distortions from our individual quasar exposures. To this
end we applied a $-$250\,\ms\ velocity shift to the wavelengths of all
echelle order centres for all our quasar exposures. The shift was
reduced linearly with distance from the order centres to reach
$+$250\,\ms\ at the order edges. The spectra were recombined again to
form a final 1-dimensional spectrum to which we fit the 4-component
model as before. The value of $\Delta\mu/\mu$ derived from this
`corrected' spectrum was $(+3.7\pm5.5)\times10^{-6}$. Compared with
the fiducial 4-component result of $(+5.6\pm5.3)\times10^{-6}$, this
represents a smaller effect, approximately $\pm1.9\times10^{-6}$, than
expected from the cruder estimates above. Given that
\citet{GriestK_10a} observed smaller distortions at shorter
wavelengths, it may also overestimate the real effect.

To summarise the above considerations, if the intra-order distortions
of the wavelength scale identified by \citet{GriestK_10a} at
5000--6500\,\AA\ also affect the much bluer wavelengths of interest
here, the implied systematic error on $\Delta\mu/\mu$ is approximately
$\pm1.9\times10^{-6}$ in our particular spectrum. However, the main
assumption in this estimate is that we have modelled the
\citeauthor{GriestK_10a} intra-order distortions reliably. The
physical explanation for those distortions is not yet known, so we do
not know whether, for example, the phase of the saw-tooth distortion
pattern with respect to the order edges is the same for all quasar
exposures taken at different times and with different telescope and
observational conditions (e.g.~telescope pointing direction, seeing
etc.), nor do we know whether the intra-order distortions follow a
general saw-tooth pattern at all, or if the distortions in the UV can
be predicted based on those in the iodine cell's calibration range.
These important questions should be resolved with future observations
and careful re-calibrations of HIRES. Thus, our estimate of
$\pm1.9\times10^{-6}$ for the systematic error due to intra-order
distortions may also be subject to model errors.

\subsubsection{General, unknown wavelength calibration
  errors}\label{sssec:wcal_general}

In Fig.~\ref{fig:K} one observes a general decrease in the $K$
coefficients for the fitted Lyman H$_2$ transitions with increasing
wavelength. A long range, monotonic distortion of the wavelength scale
is therefore nearly degenerate (though not completely) with the effect
of a shift in $\mu$ on the Lyman lines. Put another way, a systematic
effect which produced a long-range wavelength distortion would cause a
strong systematic effect on $\Delta\mu/\mu$ if only the Lyman H$_2$
transitions were fitted. It is therefore important to fit the Werner
transitions as well: their $K$ values are very different to those of
the Lyman transitions at similar wavelengths, potentially breaking the
degeneracy between $\Delta\mu/\mu$ and long-range wavelength
distortions and improving resistance to systematic errors. If we
remove the $\mu$-dependence of the 12 H$_2$ and 1 HD Werner
transitions in our fit, a 4-component model is statistically preferred
and returns $\Delta\mu/\mu=(+5.2\pm5.6)\times10^{-6}$, very similar to
the fiducial result in equation (\ref{eq:raw_result}).

However, the \SNR\ at bluer wavelengths where the Werner transitions
fall is much lower than for the reddest Lyman transitions, so the
effect of the Werner transitions in breaking this degeneracy between
$\Delta\mu/\mu$ and long-range wavelength distortions is reduced. To
illustrate this we inserted such a distortion into a simulated version
of our spectrum. The wavelength scale for one of the Monte Carlo
realizations from Appendix A was compressed according to
\begin{equation}\label{eq:comp}
\frac{\Delta v_j}{c} = \left(\frac{\Delta \mu}{\mu}\right)_{\rm sys}\left(a\lambda_j+b\right)\,,
\end{equation}
where $\lambda_j$ is the initial (observed) wavelength of a given
pixel, $\Delta v_j$ is the velocity shift applied to it,
$(\Delta\mu/\mu)_{\rm sys}$ is the systematic error in $\Delta\mu/\mu$
we seek to mimic, and the constants $a$ and $b$ are set to mimic the
general decrease in $K$ with increasing wavelength for the Lyman
transitions (see Fig.~\ref{fig:K}). To obtain a spurious shift in
$\Delta\mu/\mu$ of approximately $+15\times10^{-6}$, we set
$(\Delta\mu/\mu)_{\rm sys}=15\times10^{-6}$,
$a=-1.3\times10^{-4}$\,\AA$^{-1}$ and $b=0.4345$. Indeed, when this
spectrum is fitted with the fiducial model used to generate it,
$\Delta\mu/\mu$ shifts from $4.0\times10^{-6}$ before compression to
$16.2\times10^{-6}$ after compression, as expected. And when the
$\mu$-dependence of the 12 H$_2$ and 1 HD Werner transitions is
removed when fitting the compressed spectrum, $\Delta\mu/\mu$ only
changes by small amount to $17.3\times10^{-6}$. That is, simply
removing the Werner transitions' dependence on $\mu$ is not a very
effective test for long-range distortions of the wavelength scale.

A more meaningful test is to fit only the Lyman and Werner transitions
in the bluest part of the spectrum and not the redder Lyman
transitions. If we fit the 51 Werner and Lyman lines bluewards of
$\lambda_{\rm obs}=3230$\,\AA\ in our simulated spectrum,
$\Delta\mu/\mu$ only shifts from $6.1$ to $7.4\times10^{-6}$ when the
compression is introduced, indicating the expected resistance to this
systematic error. With the $\mu$-dependence of the 12 H$_2$ and 1 HD
Werner transitions removed in this fit, $\Delta\mu/\mu$ is $7.6$ and
$11.3\times10^{-6}$ before and after compression, respectively. That
is, if there is a long-range, monotonic distortion of the wavelength
scale, we expect to find a substantial shift in $\Delta\mu/\mu$ when
the $\mu$-dependence of the Werner transitions is removed. For the
real spectrum, the value of $\Delta\mu/\mu$ derived from the Lyman and
Werner lines bluewards of $\lambda_{\rm obs}=3230$\,\AA\ is
$(+12.1\pm11.7)\times10^{-6}$ and, when the $\mu$-dependence of the
Werner transitions is removed, this changes to
$(+12.8\pm14.3)\times10^{-6}$. Thus, despite this being a reasonable
test for long-range distortions of the wavelength scale, it fails to
reveal evidence for such an effect in our spectrum of J2123$-$0050.

\subsubsection{Errors from spectral
  re-dispersion}\label{sssec:redisperse}

As discussed in Section \ref{ssec:keck}, the spectrum to which we fit
our absorption profile models is the weighted mean of several
individual quasar exposures. In combining these spectra, they were all
re-dispersed onto the same wavelength grid. Note that the data
reduction procedures we employed meant that each exposure was only
re-binned in this way just once. Nevertheless, this rebinning
introduces correlations between the flux (and flux uncertainty) of
neighbouring pixels. In principle, this should have a small effect on
the fitted centroid of any spectral feature and so, even though we fit
many molecular lines simultaneously, a residual effect on
$\Delta\mu/\mu$ may exist. To test the possible size of this effect we
re-combined the quasar exposures onto a variety of slightly different
wavelength grids with different dispersions (i.e.~\kms\ per pixel) and
wavelength zero points. The values of $\Delta\mu/\mu$ derived from
these different versions of the final combined spectrum varied by
$\pm0.8\times10^{-6}$ at most.

\subsubsection{Velocity structure}\label{sssec:velocity}

The fiducial absorption model assumes that the molecular velocity
structure is the same in all $J$-levels (see Section
\ref{sssec:params}). In a Galactic line of sight with H$_2$
absorption, \citet{JenkinsE_97a} found that (part of) the absorption
profile becomes broader and systematically shifts in velocity with
increasing $J$. The absorption profiles comprise many velocity
components spanning velocities greater than the broadening and
shifting observed. This is consistent with the velocity structure
being the same in all $J$-levels where some constituent VCs have low
optical depths in the low-$J$ transitions but are relatively strong in
the higher-$J$ transitions. Thus, if one knew the velocity structure
-- or one used as many VCs as required to model all the statistically
significant structure in the observed line profiles -- one may not
necessarily find that individual VCs broaden or shift with increasing
$J$. Instead, one would simply find that the relative column densities
of neighbouring VCs changed with increasing $J$. However, if any
significant unmodelled structure remains, and its effective optical
depth varies as a function of $J$, it may cause a systematic effect in
$\Delta\mu/\mu$. An example of this effect in an H$_2$-bearing quasar
absorber was identified by \citet{NoterdaemeP_07b}.

\citet{IvanchikA_05a} considered this possibility, pointing out that
the potential effect on $\Delta\mu/\mu$ is diminished because
transitions of different $J$ are interspersed in wavelength space.
This is also evident in Fig.~\ref{fig:K} (lower panel). Although we
emphasise that Figs.~\ref{fig:comps} and \ref{fig:crs} demonstrate
that all the statistically significant structure in the molecular
transitions has been modelled, we can make a direct test for this
effect by allowing $\Delta\mu/\mu$ \emph{and} the velocity structure
to be different for different H$_2$ $J$-levels and HD. In Section
\ref{ssec:consistency} we made only the former assumption and found
that a stringent test, with relatively small statistical
uncertainties, was only possible when grouping together different
H$_2$ $J$-levels and HD. By introducing more free parameters in this
new test -- the different VC redshifts -- for different H$_2$
$J$-levels and HD (or groups thereof), we expect even larger
uncertainties. So we again form groups of H$_2$ $J$-levels and HD,
deriving a single value of $\Delta\mu/\mu$ with a 4-component velocity
structure in the HD and H$_2$ $J$=0--1 levels and another value with a
different 4-component model for the H$_2$ $J$=2--5 levels. The two
values are shown in Fig.~\ref{fig:J} (right panel): we see no strong
evidence for the effects of unmodelled velocity structure as a
function of species (H$_2$ $J$-level or HD).

\subsubsection{Summary of systematic errors}\label{sssec:sys_sum}

To summarise the above discussion, by far the two most important
potential systematic effects are small distortions in the ThAr
wavelength scale and intra-order distortions of the quasar wavelength
scale relative to the ThAr exposures. These contribute systematic
errors in $\Delta\mu/\mu$ of approximately $\pm2.0$ and
$\pm1.9\times10^{-6}$, respectively. Effects from re-dispersion of the
spectra may contribute as much as $\pm0.8\times10^{-6}$.

\section{Summary and discussion}\label{sec:discussion}

Our final result for the $z_{\rm abs}=2.059$ absorber observed in the
Keck spectrum of J2123$-$0050 is
\begin{equation}\label{eq:final_result}
\Delta\mu/\mu = (+5.6\pm5.5_{\rm stat}\pm2.9_{\rm sys})\times10^{-6}\,.
\end{equation}
This includes the 1-$\sigma$ statistical error [see equation
(\ref{eq:raw_result})] and the quadrature addition of the three main
potential systematic errors discussed in Section
\ref{ssec:systematics}: long- and short-range wavelength calibration
errors and effects from re-dispersion of the spectra. It is difficult
to estimate the confidence level represented by the quoted systematic
error component, though it is likely to represent a higher confidence
(i.e.~$>$68\,per cent) than the 1-$\sigma$ level represented by the
quoted statistical uncertainty. However, we emphasise that the
systematic error estimate for short-range calibration errors is based
only on the limited information about intra-order wavelength
distortions currently accessible \citep{GriestK_10a}; it is therefore
model-dependent and may under- or over-estimate the true effect, if
present.

Our result in equation (\ref{eq:final_result}) is based on fitting an
absorption profile model to particular regions of the spectrum which
contain 86 H$_2$ and 7 HD transitions, most of which are blended to
varying extents with broader Lyman-$\alpha$ forest lines, and a few of
which blend with well-constrained, weak metal absorption. All these
lines, together with local continua and zero flux level adjustments
for most lines, are fitted simultaneously to determine a single,
best-fit value of $\Delta\mu/\mu$. To reduce the number of free
parameters in the fit, two main physically motivated assumptions were
made about the molecular transitions: that the velocity structure is
the same for all $J$-levels and that individual velocity components
have the same Doppler broadening parameter in all $J$-levels. Also,
since the HD lines are so weak, it was convenient to assume that the
HD ($J=0$) velocity components follow the same relative optical depth
pattern as the H$_2$ $J=0$ transitions. Finally, the oscillator
strengths of some H$_2$ transitions were effectively treated as free
parameters because their optical depths are higher than expected based
on our fiducial model; see Section \ref{sssec:params}. However, we
have individually relaxed each of these assumptions and/or removed the
transitions they pertain to, and find the fiducial result in equation
(\ref{eq:final_result}) to be quite insensitive to them.

\begin{figure}
\begin{center}
\includegraphics[width=\columnwidth]{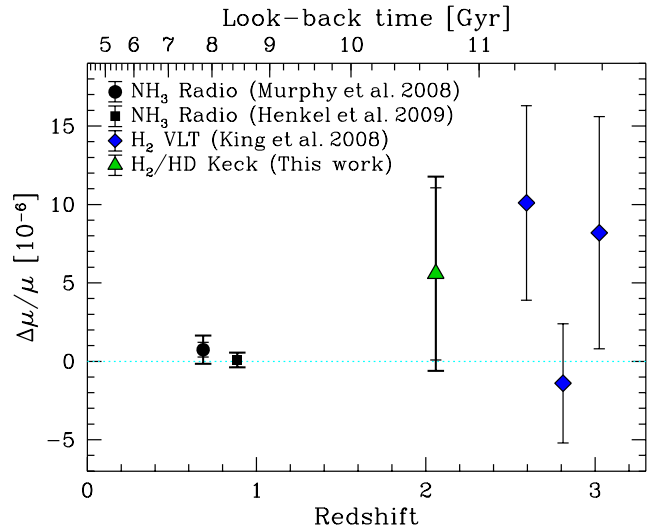}\vspace{-0.5em}
\caption{Current extragalactic $\Delta\mu/\mu$ constraints. The legend
  gives the reference for each point and the method used, i.e.~H$_2$
  or NH$_3$ quasar absorption lines. 1-$\sigma$ statistical
  uncertainties are indicated with thinner error bars with shorter
  terminators while the thicker error bars with longer terminators
  show the systematic error added in quadrature with the statistical
  ones. Diamonds represent the VLT constraints; our new measurement is
  the only one from Keck. Note that two of the \citet{KingJ_08a}
  absorbers use the same raw VLT data as those analysed previously by
  \citet{IvanchikA_05a} and \citet{ReinholdE_06a}. We omit the results
  of those two studies assuming that the improved analysis in
  \citeauthor{KingJ_08a} has yielded more reliable results. For
  clarity, we also omit the other recent re-analyses of the same two
  absorbers by \citet{WendtM_08a} and \citet{ThompsonR_09a}; these
  works also yielded null results with somewhat larger errors than
  reported by \citeauthor{KingJ_08a}.}
\label{fig:summary}
\end{center}
\end{figure}

Figure \ref{fig:summary} compares the contemporary astronomical
$\Delta\mu/\mu$ constraints outside our Galaxy. Our new constraint is
the first from Keck with precision comparable to those from the VLT.
Agreement is seen between our Keck result and the three VLT
constraints from \citet{KingJ_08a} who used similar calibration and
fitting techniques to those described here. Such consistency is
encouraging, allowing us to (at least formally) derive a weighted
mean\footnote{We added the statistical and systematic error components
  on our new result in quadrature for this calculation. No systematic
  error estimates are available for the \citeauthor{KingJ_08a}
  results.} value of $\Delta\mu/\mu=(+3.5\pm2.8)\times10^{-6}$ for
redshifts $z=2.0$--$3.1$.  Statistical errors are shown as thinner
error bars with shorter terminators in Fig.~\ref{fig:summary}.
Although the statistical error in our new measurement is slightly
smaller than for two of the \citet{KingJ_08a} absorbers, it is
slightly worse than for the $z_{\rm abs}=2.811$ absorber towards
Q\,0528$-$250, i.e.~$3.9\times10^{-6}$.  The main reason for this is
the lower \SNR\ of our current Keck spectrum of J2123$-$0050, despite
the fact that its spectral resolution is higher and more molecular
transitions were utilized in our analysis.  Since J2123$-$0050 is
relatively bright, improvements in both the statistical and systematic
uncertainties can be achieved with new observations of moderate
duration. And since it is located near the celestial equator,
J2123$-$0050 can easily be observed from both Keck and VLT, enabling a
further opportunity to check for instrumental systematic errors. We
have recently undertaken such observations and a future paper will
report the results.

At $z<1$, comparison of the radio inversion transitions of NH$_3$ --
which have enhanced sensitivity to $\mu$-variation
\citep{VeldhovenJ_04a,FlambaumV_07a} -- with less sensitive molecular
rotational lines (e.g.~HCO$^+$, HCN), has yielded two very strong
constraints, $\Delta\mu/\mu=(+0.74\pm0.47_{\rm stat}\pm0.76_{\rm
  sys})\times10^{-6}$ at $z=0.685$ \citep{MurphyM_08b} and
$(+0.08\pm0.47_{\rm sys})\times10^{-6}$ at $z=0.889$
\citep{HenkelC_09a}. These results are plotted in
Fig.~\ref{fig:summary} for comparison with the higher redshift
H$_2$/HD constraints. It is clear that the radio constraints have
superior precision and, by current estimates, smaller potential
systematic errors. However, direct comparison of the radio and optical
constraints is difficult because of the possibility, in principle, for
spatial variations in $\mu$: the different molecular species (NH$_3$
and H$_2$/HD) trace regions of different densities and, therefore,
different spatial scales and environment. If $\mu$ does vary, we do
not know what that variation depends on, so it is presumptuous to
prefer one type of measurement over the other. Indeed, this is
highlighted by \citet{LevshakovS_09a} who studied NH$_3$ inversion
\emph{emission} lines from numerous Galactic molecular clouds. With
statistical errors from previous literature of $\sim$0.1\,\kms\ they
find velocity offsets between the NH$_3$ inversion and rotational
molecular emission of up to $\left|\Delta v\right|\sim0.5$\,\kms\ in
individual systems. This might indicate spatial variations in $\mu$
throughout our Galaxy, although intrinsic shifts between emission
lines of different molecules are to be expected.  The possibility for
both space- and time-variations in $\mu$ is even more important given
the different redshift ranges currently probed by the radio and
optical constraints.

Several other quasar absorption- and emission-line techniques for
constraining variations in combinations of fundamental constants
involving $\mu$ are also of note. For example, comparison of H{\sc
  \,i} 21-cm absorption with metal absorption lines constrains the
quantity $X\equiv g_{\rm p}\alpha^2/\mu$ where $g_{\rm p}$ is the
proton $g$-factor. A variety of analyses using different metal-ions
have been performed
\citep[e.g.][]{WolfeA_76a,TzanavarisP_05a,KanekarN_06a}, with
measurements from 9 absorption systems at $0.23<z_{\rm abs}<2.35$ by
\citet{TzanavarisP_07a} providing a constraint of $\Delta
X/X=(+6.3\pm9.9)\times10^{-6}$. However, the fact that the H{\sc \,i}
21-cm and metal-line velocity structures are observationally dissimilar
-- probably because the radio morphology of most background quasars is
not point-like -- means that improvements must come by averaging over
many sight-lines and/or carefully selecting quasars with point-like
radio morphologies. Comparing H{\sc \,i} 21-cm with the `main' OH
18-cm absorption lines, which constrains $F\equiv g_{\rm
  p}(\mu\alpha^2)^{1.57}$ \citep{ChengalurJ_03a}, suffers less from
this problem because of the similarity of the wavelengths concerned.
\citet{KanekarN_05a} analysed two absorbers at $z_{\rm abs}=0.685$ and
$0.765$ to obtain the constraint $\Delta F/F=(+4.4\pm3.6_{\rm
  stat}\pm10_{\rm sys})\times10^{-6}$. Finally, a promising technique
is to use the `satellite' OH 18-cm lines which, when the ground state
level populations are inverted, results in the two transitions being
`conjugate' -- one occurs in emission, the other in absorption, but
they have the same optical depth profile and are guaranteed to arise
in the same gas clouds. Comparing the sum and difference of the
optical depth profiles of the 1612 and 1720\,MHz conjugate satellite
lines constrains $G\equiv g_{\rm p}(\mu\alpha^2)^{1.85}$
\citep{ChengalurJ_03a}. With only two such systems known outside our
Galaxy, and with their published spectra having very low \SNR, no
definitive constraints on $\Delta G/G$ yet exist
\citep{DarlingJ_04a,KanekarN_05a}.  \citet{KanekarN_08a} has presented
a preliminary constraint of $\left|\Delta G/G\right|<11\times10^{-6}$
for a system at $z=0.247$ and also $<12\times10^{-6}$ for a Galactic
system, Centaurus A.

While these constraints on variations in $X$, $F$ and $G$ are
important in their own right, it is difficult to compare them with
constraints on $\Delta\mu/\mu$. As we mentioned in Section
\ref{sec:intro}, predictions for the relationships between variations
in, e.g., $\mu$ and $\alpha$ are strongly model-dependent
\citep{DentT_08a}. Currently, there is also no overlap with the
redshift range occupied by the H$_2$/HD constraints on $\Delta\mu/\mu$
and variations in $X$, $F$ and/or $G$.

A striking feature of Fig.~\ref{fig:summary} is the paucity of direct
$\mu$ measurements outside our Galaxy. Compare this with, for example,
the 143 constraints on $\alpha$-variation available from Keck/HIRES
over the redshift range $z=0.2$--$4.2$ \citep{MurphyM_04a}. Clearly,
much larger samples of both NH$_3$ and H$_2$/HD constraints, in
overlapping redshift ranges, would allow additional tests for
systematic errors and stronger conclusions to be drawn. In this
respect it is important to reduce the statistical errors in individual
H$_2$/HD measurements. This can only be done by substantially
increasing the SNR of the optical spectra. We have demonstrated here
that, at least for the Keck/HIRES spectrum of J2123$-$0050, the
statistical error still dominates over systematic errors [equation
(\ref{eq:final_result})]. Nevertheless, if such systematic errors are
not also reduced by improving the spectral SNR, it becomes more
important to greatly increase the number of H$_2$/HD absorbers for
measuring $\Delta\mu/\mu$ in order to approach the precision
demonstrated in the NH$_3$ measurements. Surveys targeting
particularly gas- and/or metal-rich absorbers have been successful in
discovering more H$_2$/HD systems
\citep[e.g.][]{LedouxC_03a,PetitjeanP_06a,NoterdaemeP_08a}. The large
number of damped Lyman-$\alpha$ systems identified in the Sloan
Digital Sky Survey \citep[e.g.][]{ProchaskaJ_09a} offers a new avenue for
discovery, J2123$-$0050 being one such example. Of course, finding
more NH$_3$ absorbers, particularly at $z>1$ would be highly desirable
as well.

\section*{Acknowledgments}

The data presented herein were obtained at the W.~M.~Keck Observatory,
which is operated as a scientific partnership among the California
Institute of Technology, the University of California and the National
Aeronautics and Space Administration. The Observatory was made
possible by the generous financial support of the W.~M.~Keck
Foundation. The authors wish to recognise and acknowledge the very
significant cultural role and reverence that the summit of Mauna Kea
has always had within the indigenous Hawaiian community. We are most
fortunate to have the opportunity to conduct observations from this
mountain. We thank the anonymous referee for their measured and
insightful review and for helping us improve the manuscript. MTM
thanks the Australian Research Council for a QEII Research Fellowship
(DP0877998). JXP is supported by NSF grants AST-0709235 and
AST-0548180. WU acknowledges financial support from the Netherlands
Foundation for Fundamental Research of Matter (FOM).


\begin{thebibliography}{}
\small
\itemindent -0.48cm

\bibitem[\protect\citeauthoryear{{Abgrall} \& {Roueff}}{{Abgrall} \&  {Roueff}}{2006}]{AbgrallH_06a}{Abgrall} H.,  {Roueff} E.,  2006, \aap, 445, 361

\bibitem[\protect\citeauthoryear{{Abgrall}, {Roueff} \& {Drira}}{{Abgrall}  et~al.}{2000}]{AbgrallH_00a}{Abgrall} H.,  {Roueff} E.,    {Drira} I.,  2000, \aaps, 141, 297

\bibitem[\protect\citeauthoryear{{Abgrall}, {Roueff}, {Launay} \& {Roncin}}{Abgrall et~al.}{1994}]{AbgrallH_94a}{Abgrall} H.,  {Roueff} E.,  {Launay} F.,    {Roncin} J.-Y.,  1994, \cjp, 72,  856

\bibitem[\protect\citeauthoryear{{Abgrall}, {Roueff}, {Launay}, {Roncin} \& {Subtil}}{Abgrall et~al.}{1993}]{AbgrallH_93c}{Abgrall} H.,  {Roueff} E.,  {Launay} F.,  {Roncin} J.Y.,    {Subtil} J.L.,  1993, \jms, 157, 512

\bibitem[\protect\citeauthoryear{{Bahcall}, {Sargent} \& {Schmidt}}{{Bahcall}  et~al.}{1967}]{BahcallJ_67b}{Bahcall} J.N.,  {Sargent} W.L.W.,    {Schmidt} M.,  1967, \apjl, 149, L11

\bibitem[\protect\citeauthoryear{{Bailly}, {Salumbides}, {Vervloet} \& {Ubachs}}{Bailly et~al.}{2009}]{BaillyD_09a}{Bailly} D.,  {Salumbides} E.J.,  {Vervloet} M.,    {Ubachs} W.,  2009, \mp,  accepted, DOI 10.1080/00268970903413350

\bibitem[\protect\citeauthoryear{{Blatt}, {Ludlow}, {Campbell}, {Thomsen}, {Zelevinsky}, {Boyd}, {Ye}, {Baillard}, {Fouch{\'e}}, {Le Targat}, {Brusch}, {Lemonde}, {Takamoto}, {Hong}, {Katori} \& {Flambaum}}{Blatt et~al.}{2008}]{BlattS_08a}{Blatt} S.~{et al.},  2008, \prl, 100, 140801

\bibitem[\protect\citeauthoryear{{Calmet} \& {Fritzsch}}{{Calmet} \&  {Fritzsch}}{2002}]{CalmetX_02a}{Calmet} X.,  {Fritzsch} H.,  2002, \epjc, 24, 639

\bibitem[\protect\citeauthoryear{{Chengalur} \& {Kanekar}}{{Chengalur} \&  {Kanekar}}{2003}]{ChengalurJ_03a}{Chengalur} J.N.,  {Kanekar} N.,  2003, \prl, 91, 241302

\bibitem[\protect\citeauthoryear{{Cing{\"o}z}, {Lapierre}, {Nguyen}, {Leefer}, {Budker}, {Lamoreaux} \& {Torgerson}}{Cing{\"o}z et~al.}{2007}]{CingozA_07a}{Cing{\"o}z} A.,  {Lapierre} A.,  {Nguyen} A.-T.,  {Leefer} N.,  {Budker} D.,  {Lamoreaux} S.K.,    {Torgerson} J.R.,  2007, \prl, 98, 040801

\bibitem[\protect\citeauthoryear{{Cowie} \& {Songaila}}{{Cowie} \&  {Songaila}}{1995}]{CowieL_95a}{Cowie} L.L.,  {Songaila} A.,  1995, \apj, 453, 596

\bibitem[\protect\citeauthoryear{{Darling}}{{Darling}}{2004}]{DarlingJ_04a}{Darling} J.,  2004, \apj, 612, 58

\bibitem[\protect\citeauthoryear{{Dent}, {Stern} \& {Wetterich}}{{Dent}  et~al.}{2008}]{DentT_08a}{Dent} T.,  {Stern} S.,    {Wetterich} C.,  2008, \prd, 78, 103518

\bibitem[\protect\citeauthoryear{{Edlen}}{{Edlen}}{1966}]{EdlenB_66a}{Edlen} B.,  1966, \met, 2, 71

\bibitem[\protect\citeauthoryear{{Flambaum} \& {Kozlov}}{{Flambaum} \&  {Kozlov}}{2007}]{FlambaumV_07a}{Flambaum} V.V.,  {Kozlov} M.G.,  2007, \prl, 98, 240801

\bibitem[\protect\citeauthoryear{{Griest}, {Whitmore}, {Wolfe}, {Prochaska}, {Howk} \& {Marcy}}{Griest et~al.}{2010}]{GriestK_10a}{Griest} K.,  {Whitmore} J.B.,  {Wolfe} A.M.,  {Prochaska} J.X.,  {Howk}  J.~C.,    {Marcy} G.~W.,  2010, \apj, 708, 158

\bibitem[\protect\citeauthoryear{{Henkel}, {Menten}, {Murphy}, {Jethava}, {Flambaum}, {Braatz}, {Muller}, {Ott} \& {Mao}}{Henkel et~al.}{2009}]{HenkelC_09a}{Henkel} C.~{et al.},  2009, \aap,  500, 725

\bibitem[\protect\citeauthoryear{{Hinnen}, {Hogervorst}, {Stolte} \& {Ubachs}}{Hinnen et~al.}{1994}]{HinnenP_94a}{Hinnen} P.C.,  {Hogervorst} W.,  {Stolte} S.,    {Ubachs} W.,  1994, Canadian  Journal of Physics, 72, 1032

\bibitem[\protect\citeauthoryear{{Hollenstein}, {Reinhold}, {de Lange} \& {Ubachs}}{Hollenstein et~al.}{2006}]{HollensteinU_06a}{Hollenstein} U.,  {Reinhold} E.,  {de Lange} C.A.,    {Ubachs} W.,  2006,  \jpb, 39, L195

\bibitem[\protect\citeauthoryear{{Ivanchik}, {Petitjean}, {Varshalovich}, {Aracil}, {Srianand}, {Chand}, {Ledoux} \& {Boiss{\'e}}}{Ivanchik et~al.}{2005}]{IvanchikA_05a}{Ivanchik} A.,  {Petitjean} P.,  {Varshalovich} D.,  {Aracil} B.,  {Srianand}  R.,  {Chand} H.,  {Ledoux} C.,    {Boiss{\'e}} P.,  2005, \aap, 440, 45

\bibitem[\protect\citeauthoryear{{Ivanov}, {Roudjane}, {Vieitez}, {de Lange}, {Tchang-Brillet} \& {Ubachs}}{Ivanov et~al.}{2008}]{IvanovT_08a}{Ivanov} T.I.,  {Roudjane} M.,  {Vieitez} M.O.,  {de Lange} C.A.,  {Tchang-Brillet} W.-{\"U}.~L.,    {Ubachs} W.,  2008, \prl, 100, 093007

\bibitem[\protect\citeauthoryear{{Jenkins} \& {Peimbert}}{{Jenkins} \&  {Peimbert}}{1997}]{JenkinsE_97a}{Jenkins} E.B.,  {Peimbert} A.,  1997, \apj, 477, 265

\bibitem[\protect\citeauthoryear{{Jennings}, {Bragg} \& {Brault}}{{Jennings}  et~al.}{1984}]{JenningsD_84a}{Jennings} D.E.,  {Bragg} S.L.,    {Brault} J.W.,  1984, \apjl, 282, L85

\bibitem[\protect\citeauthoryear{{Kanekar}}{{Kanekar}}{2008}]{KanekarN_08a}{Kanekar} N.,  2008, \mpla, 23, 2711

\bibitem[\protect\citeauthoryear{{Kanekar}, {Carilli}, {Langston}, {Rocha}, {Combes}, {Subrahmanyan}, {Stocke}, {Menten}, {Briggs} \& {Wiklind}}{Kanekar et~al.}{2005}]{KanekarN_05a}{Kanekar} N.~{et al.},  2005, \prl, 95, 261301

\bibitem[\protect\citeauthoryear{{Kanekar}, {Subrahmanyan}, {Ellison}, {Lane} \& {Chengalur}}{Kanekar et~al.}{2006}]{KanekarN_06a}{Kanekar} N.,  {Subrahmanyan} R.,  {Ellison} S.L.,  {Lane} W.M.,  {Chengalur} J.N.,  2006, \mnras, 370, L46

\bibitem[\protect\citeauthoryear{{King}, {Webb}, {Murphy} \& {Carswell}}{{King}  et~al.}{2008}]{KingJ_08a}{King} J.A.,  {Webb} J.K.,  {Murphy} M.T.,    {Carswell} R.~F.,  2008, \prl,  101, 251304

\bibitem[\protect\citeauthoryear{{Ledoux}, {Petitjean} \& {Srianand}}{{Ledoux}  et~al.}{2003}]{LedouxC_03a}{Ledoux} C.,  {Petitjean} P.,    {Srianand} R.,  2003, \mnras, 346, 209

\bibitem[\protect\citeauthoryear{{Levshakov}, {Molaro} \& {Kozlov}}{{Levshakov}  et~al.}{2008}]{LevshakovS_09a}{Levshakov} S.A.,  {Molaro} P.,    {Kozlov} M.G.,  2008, preprint  (arXiv:0808.0583)

\bibitem[\protect\citeauthoryear{{Liddle}}{{Liddle}}{2007}]{LiddleA_07a}{Liddle} A.R.,  2007, \mnras, 377, L74

\bibitem[\protect\citeauthoryear{{Marion}, {Santos}, {Abgrall}, {Zhang}, {Sortais}, {Bize}, {Maksimovic}, {Calonico}, {Gruenert}, {Mandache}, {Lemonde}, {Santarelli}, {Laurent}, {Clairon} \& {Salomon}}{Marion et~al.}{2003}]{MarionH_03a}{Marion} H.~{et al.},  2003, \prl, 90, 150801

\bibitem[\protect\citeauthoryear{{Meshkov}, {Stolyarov}, {Ivanchik} \& {Varshalovich}}{Meshkov et~al.}{2006}]{MeshkovV_06a}{Meshkov} V.V.,  {Stolyarov} A.V.,  {Ivanchik} A.V.,    {Varshalovich}  D.~A.,  2006, \spjetpl, 83, 303

\bibitem[\protect\citeauthoryear{{Murphy}, {Flambaum}, {Muller} \& {Henkel}}{Murphy et~al.}{2008}]{MurphyM_08b}{Murphy} M.T.,  {Flambaum} V.V.,  {Muller} S.,    {Henkel} C.,  2008, \sci,  320, 1611

\bibitem[\protect\citeauthoryear{{Murphy}, {Flambaum}, {Webb}, {Dzuba}, {Prochaska} \& {Wolfe}}{Murphy et~al.}{2004}]{MurphyM_04a}{Murphy} M.T.,  {Flambaum} V.V.,  {Webb} J.K.,  {Dzuba} V.~V.,  {Prochaska}  J.~X.,    {Wolfe} A.~M.,  2004, \lnp, 648, 131

\bibitem[\protect\citeauthoryear{{Murphy}, {Tzanavaris}, {Webb} \& {Lovis}}{Murphy et~al.}{2007}]{MurphyM_07b}{Murphy} M.T.,  {Tzanavaris} P.,  {Webb} J.K.,    {Lovis} C.,  2007, \mnras,  378, 221

\bibitem[\protect\citeauthoryear{{Murphy}, {Webb} \& {Flambaum}}{{Murphy}  et~al.}{2003}]{MurphyM_03a}{Murphy} M.T.,  {Webb} J.K.,    {Flambaum} V.V.,  2003, \mnras, 345, 609

\bibitem[\protect\citeauthoryear{{Murphy}, {Webb} \& {Flambaum}}{{Murphy}  et~al.}{2008}]{MurphyM_08a}{Murphy} M.T.,  {Webb} J.K.,    {Flambaum} V.V.,  2008, \mnras, 384, 1053

\bibitem[\protect\citeauthoryear{{Noterdaeme}, {Ledoux}, {Petitjean}, {Le Petit}, {Srianand} \& {Smette}}{Noterdaeme et~al.}{2007}]{NoterdaemeP_07b}{Noterdaeme} P.,  {Ledoux} C.,  {Petitjean} P.,  {Le Petit} F.,  {Srianand} R.,     {Smette} A.,  2007, \aap, 474, 393

\bibitem[\protect\citeauthoryear{{Noterdaeme}, {Ledoux}, {Petitjean} \& {Srianand}}{Noterdaeme et~al.}{2008}]{NoterdaemeP_08a}{Noterdaeme} P.,  {Ledoux} C.,  {Petitjean} P.,    {Srianand} R.,  2008, \aap,  481, 327

\bibitem[\protect\citeauthoryear{{Noterdaeme}, {Petitjean}, {Ledoux}, {Srianand} \& {Ivanchik}}{Noterdaeme et~al.}{2008}]{NoterdaemeP_08b}{Noterdaeme} P.,  {Petitjean} P.,  {Ledoux} C.,  {Srianand} R.,    {Ivanchik}  A.,  2008, \aap, 491, 397

\bibitem[\protect\citeauthoryear{{Peik}, {Lipphardt}, {Schnatz}, {Schneider}, {Tamm} \& {Karshenboim}}{Peik et~al.}{2004}]{PeikE_04a}{Peik} E.,  {Lipphardt} B.,  {Schnatz} H.,  {Schneider} T.,  {Tamm} C.,  {Karshenboim} S.G.,  2004, \prl, 93, 170801

\bibitem[\protect\citeauthoryear{{Petitjean}, {Ledoux}, {Noterdaeme} \& {Srianand}}{Petitjean et~al.}{2006}]{PetitjeanP_06a}{Petitjean} P.,  {Ledoux} C.,  {Noterdaeme} P.,    {Srianand} R.,  2006, \aap,  456, L9

\bibitem[\protect\citeauthoryear{{Philip}, {Sprengers}, {Pielage}, {de Lange}, {Ubachs} \& {Reinhold}}{Philip et~al.}{2004}]{PhilipJ_04a}{Philip} J.,  {Sprengers} J.P.,  {Pielage} T.,  {de Lange} C.A.,  {Ubachs}  W.,    {Reinhold} E.,  2004, \cjc, 82, 713

\bibitem[\protect\citeauthoryear{{Prestage}, {Tjoelker} \& {Maleki}}{{Prestage}  et~al.}{1995}]{PrestageJ_95a}{Prestage} J.D.,  {Tjoelker} R.L.,    {Maleki} L.,  1995, \prl, 74, 3511

\bibitem[\protect\citeauthoryear{{Prochaska} \& {Wolfe}}{{Prochaska} \&  {Wolfe}}{2009}]{ProchaskaJ_09a}{Prochaska} J.X.,  {Wolfe} A.M.,  2009, \apj, 696, 1543

\bibitem[\protect\citeauthoryear{{Reinhold}, {Buning}, {Hollenstein}, {Ivanchik}, {Petitjean} \& {Ubachs}}{Reinhold et~al.}{2006}]{ReinholdE_06a}{Reinhold} E.,  {Buning} R.,  {Hollenstein} U.,  {Ivanchik} A.,  {Petitjean}  P.,    {Ubachs} W.,  2006, \prl, 96, 151101

\bibitem[\protect\citeauthoryear{{Rosenband}, {Hume}, {Schmidt}, {Chou}, {Brusch}, {Lorini}, {Oskay}, {Drullinger}, {Fortier}, {Stalnaker}, {Diddams}, {Swann}, {Newbury}, {Itano}, {Wineland} \& {Bergquist}}{Rosenband et~al.}{2008}]{RosenbandT_08a}{Rosenband} T.~{et al.},  2008, \sci, 319,  1808

\bibitem[\protect\citeauthoryear{{Salumbides}, {Bailly}, {Khramov}, {Wolf}, {Eikema}, {Vervloet} \& {Ubachs}}{Salumbides et~al.}{2008}]{SalumbidesE_08a}{Salumbides} E.J.,  {Bailly} D.,  {Khramov} A.,  {Wolf} A.L.,  {Eikema}  K.S.~E.,  {Vervloet} M.,    {Ubachs} W.,  2008, \prl, 101, 223001

\bibitem[\protect\citeauthoryear{{Shelkovnikov}, {Butcher}, {Chardonnet} \& {Amy-Klein}}{Shelkovnikov et~al.}{2008}]{ShelkovnikovA_08a}{Shelkovnikov} A.,  {Butcher} R.J.,  {Chardonnet} C.,    {Amy-Klein} A.,  2008, \prl, 100, 150801

\bibitem[\protect\citeauthoryear{{Thompson}}{{Thompson}}{1975}]{ThompsonR_75a}{Thompson} R.I.,  1975, \al, 16, 3

\bibitem[\protect\citeauthoryear{{Thompson}, {Bechtold}, {Black}, {Eisenstein}, {Fan}, {Kennicutt}, {Martins}, {Prochaska} \& {Shirley}}{Thompson et~al.}{2009}]{ThompsonR_09a}{Thompson} R.I.~{et al.},  2009, \apj, 703, 1648

\bibitem[\protect\citeauthoryear{{Tzanavaris}, {Murphy}, {Webb}, {Flambaum} \& {Curran}}{Tzanavaris et~al.}{2007}]{TzanavarisP_07a}{Tzanavaris} P.,  {Murphy} M.T.,  {Webb} J.K.,  {Flambaum} V.V.,    {Curran}  S.~J.,  2007, \mnras, 374, 634

\bibitem[\protect\citeauthoryear{{Tzanavaris}, {Webb}, {Murphy}, {Flambaum} \& {Curran}}{Tzanavaris et~al.}{2005}]{TzanavarisP_05a}{Tzanavaris} P.,  {Webb} J.K.,  {Murphy} M.T.,  {Flambaum} V.V.,    {Curran}  S.~J.,  2005, \prl, 95, 041301

\bibitem[\protect\citeauthoryear{{Ubachs}, {Buning}, {Eikema} \& {Reinhold}}{Ubachs et~al.}{2007}]{UbachsW_07a}{Ubachs} W.,  {Buning} R.,  {Eikema} K.S.E.,    {Reinhold} E.,  2007, \jms,  241, 155

\bibitem[\protect\citeauthoryear{{Ubachs}, {Eikema}, {Hogervorst} \& {Cacciani}}{Ubachs et~al.}{1997}]{UbachsW_97a}{Ubachs} W.,  {Eikema} K.S.E.,  {Hogervorst} W.,    {Cacciani} P.C.,  1997,  \josb, 14, 2469

\bibitem[\protect\citeauthoryear{{Ubachs} \& {Reinhold}}{{Ubachs} \&  {Reinhold}}{2004}]{UbachsW_04a}{Ubachs} W.,  {Reinhold} E.,  2004, \prl, 92, 101302

\bibitem[\protect\citeauthoryear{{van Veldhoven}, {K{\"u}pper}, {Bethlem}, {Sartakov}, {van  Roij} \& {Meijer}}{van Veldhoven et~al.}{2004}]{VeldhovenJ_04a}{van Veldhoven} J.,  {K{\"u}pper} J.,  {Bethlem} H.L.,  {Sartakov} B.,  {van  Roij} A.J.A.,    {Meijer} G.,  2004, European Physical Journal D, 31, 337

\bibitem[\protect\citeauthoryear{{Varshalovich}, {Ivanchik}, {Petitjean}, {Srianand} \& {Ledoux}}{Varshalovich et~al.}{2001}]{VarshalovichD_01a}{Varshalovich} D.A.,  {Ivanchik} A.V.,  {Petitjean} P.,  {Srianand} R.,  {Ledoux} C.,  2001, \al, 27, 683

\bibitem[\protect\citeauthoryear{{Varshalovich} \& {Levshakov}}{{Varshalovich}  \& {Levshakov}}{1993}]{VarshalovichD_93b}{Varshalovich} D.A.,  {Levshakov} S.A.,  1993, \spjetpl, 58, 237

\bibitem[\protect\citeauthoryear{{Vogt et~al.}}{{Vogt et~al.}}{1994}]  {VogtS_94a}{Vogt} S.S. {et~al.,} 1994, in {Crawford} D.L.,  {Craine} E.R.,  eds,  Proc. SPIE Vol. 2198, Instrumentation in Astronomy VIII. p.~362

\bibitem[\protect\citeauthoryear{{Wendt} \& {Reimers}}{{Wendt} \&  {Reimers}}{2008}]{WendtM_08a}{Wendt} M.,  {Reimers} D.,  2008, \epjst, 163, 197

\bibitem[\protect\citeauthoryear{{Wolfe}, {Brown} \& {Roberts}}{{Wolfe}  et~al.}{1976}]{WolfeA_76a}{Wolfe} A.M.,  {Brown} R.L.,    {Roberts} M.S.,  1976, \prl, 37, 179

\end{thebibliography}

\appendix

\section{Monte Carlo test of fitting algorithm to determine \boldmath{$\Delta\mu/\mu$}}\label{app:a}

The fiducial fit performed on the spectrum of J2123$-$0050 to
determine $\Delta\mu/\mu$ is a fairly large and complex one. It
contains more fitted spectral pixels and more fitted parameters than
most, if not all, previous fits of its kind in the literature.
Furthermore, the fitted parameters are not all free and independent,
but are tied together in a variety of physically meaningful ways.
Therefore, to ensure that the fitting code {\sc vpfit} determines
$\Delta\mu/\mu$ correctly in this context, we performed a Monte Carlo
test using simulated spectra produced from the fiducial fit we
established to the real data. That is, each simulated spectrum is
produced simply by taking the fit with 4 molecular velocity
components, plus all the fitted Lyman-$\alpha$ forest lines,
interloping metal lines, local continuum and zero flux level
adjustments, and adding Gaussian noise with a $\sigma$ of 0.8 times
the error array of the real spectrum. The artificially high \SNR\ in
the simulations was used to ensure that Lyman-$\alpha$ blends that
were only marginally statistically required in the fit to the real
spectrum were not removed statistically when fitting the simulated
versions.

\begin{figure}
\begin{center}
\includegraphics[width=0.9\columnwidth]{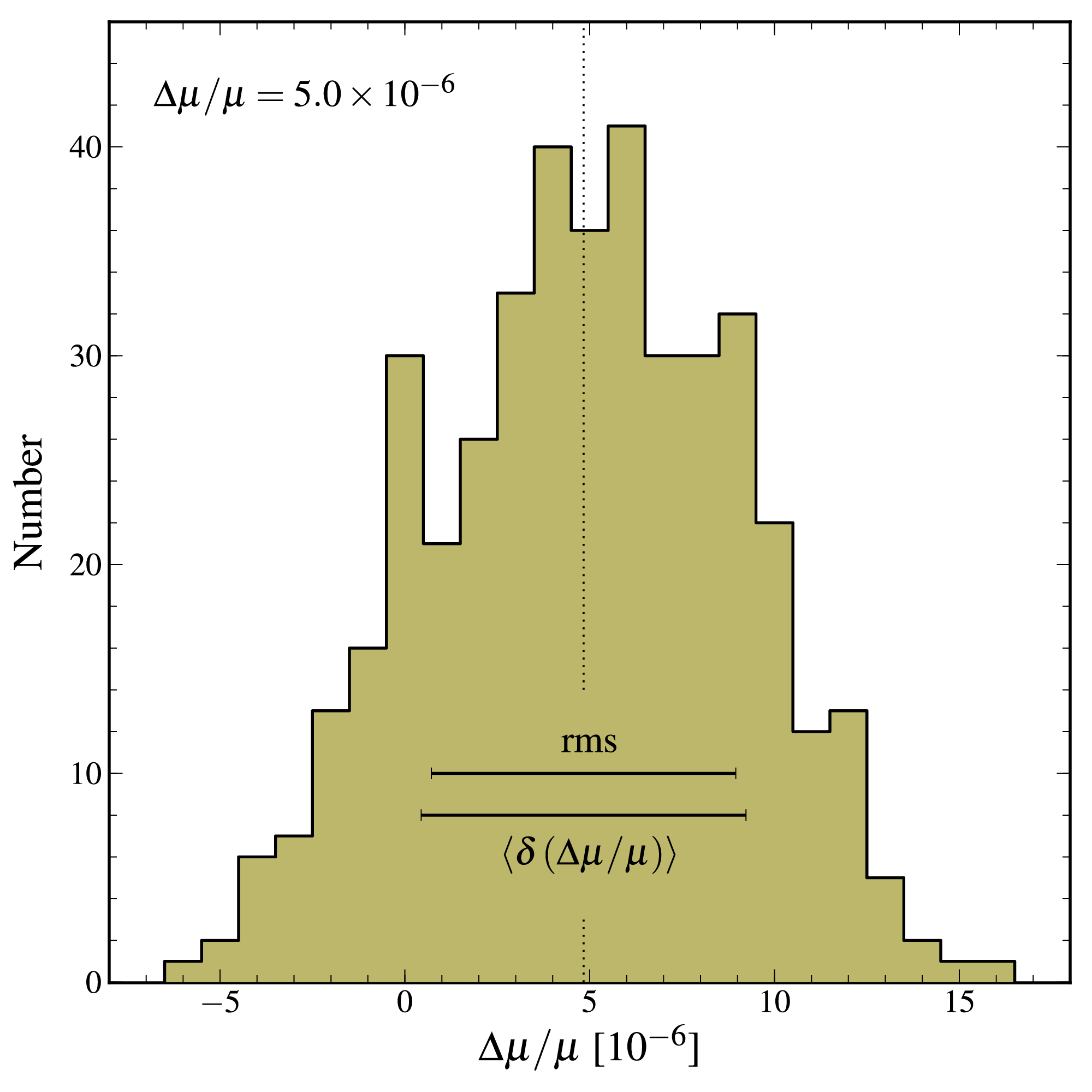}\vspace{-0.5em}
\caption{420 Monte Carlo simulations of the fiducial 4 component
  absorption model used in our analysis. The value of $\Delta\mu/\mu$
  returned from the simulated spectra (vertical dotted line)
  corresponds well to the input value, $+5\times10^{-6}$. The mean
  1-$\sigma$ error and the standard deviation are shown to be
  consistent.}
\label{fig:sims}
\end{center}
\end{figure}

Each simulated spectrum was then fitted with {\sc vpfit} with initial
guess parameters set to be those used to produce the simulated
spectrum. An input value of $\Delta\mu/\mu=+5\times10^{-6}$ was used
to test whether {\sc vpfit} recovered that value when started from an
initial guess of zero. Figure \ref{fig:sims} shows the results of
fitting 420 simulated spectra. The mean 1-$\sigma$ uncertainty derived
on individual $\Delta\mu/\mu$ measurements is $4.4\times10^{-6}$ which
is the expected value given the value of $5.5\times10^{-6}$ from the
real spectrum and a scaling of 0.8 times its flux errors. This
corresponds well to the RMS of $4.1\times10^{-6}$ for the 420
$\Delta\mu/\mu$ measurements; it may even be that the individual
uncertainties returned by {\sc vpfit} are slightly conservative. The
mean $\Delta\mu/\mu$ value retrieved is $4.8\times10^{-6}$.

\bspsmall

\label{lastpage}

\section{Supporting Information}

Figure \ref{fig:fit_all_supp} is the complete version of
Fig.~\ref{fig:fit_all}. It will appear in the online version of this
paper, not in the printed version. Table B1, the complete version of
Table 1, is provided in the online version only as a machine-readable
ASCII file.

\clearpage

\renewcommand \thefigure{B1}
\begin{figure*}
\begin{center}
\includegraphics[width=0.92\textwidth]{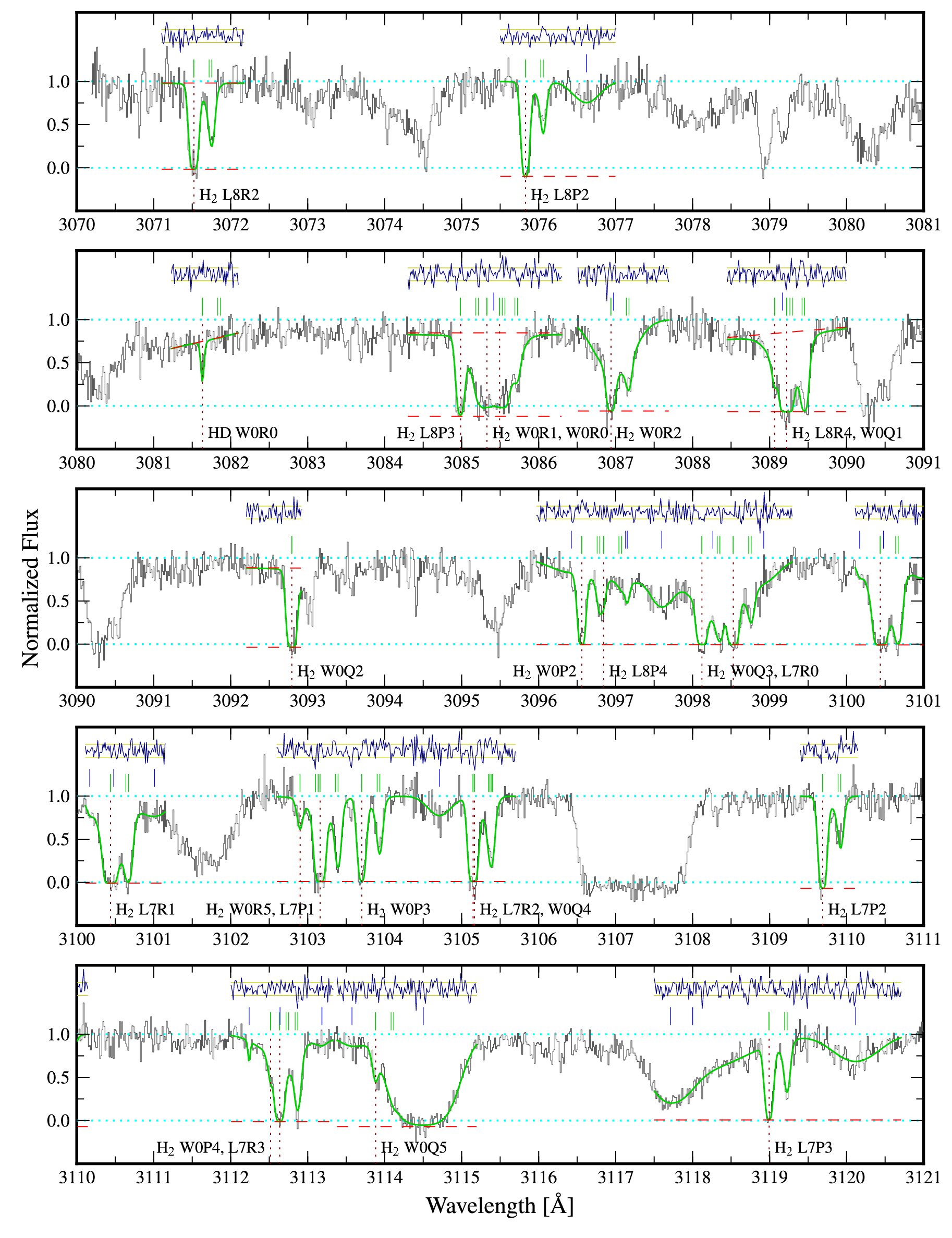}\vspace{-0.5em}
\caption{All regions of the J2123$-$0050 Keck spectrum fitted
  simultaneously in our analysis. The spectrum (black histogram) is
  normalized by a nominal continuum (upper dotted line) fitted over
  large spectral scales. Local linear continua (upper dashed lines)
  and zero levels (lower dashed lines) are fitted simultaneously with
  the H$_2$/HD and broader Lyman-$\alpha$ lines. The fits are shown
  with solid grey/green lines. H$_2$/HD transitions are labelled and
  their constituent velocity components are indicated by grey/green
  tick-marks immediately above the spectrum. Higher above the spectrum
  are tick-marks indicating the positions of Lyman-$\alpha$ lines
  (blue) and Fe{\sc \,ii} lines (red). Note that the metal-line
  velocity structure is constrained with the Fe{\sc
    \,ii}\,$\lambda$1608\,\AA\ transition shown in the final panel of
  the figure. The residual spectrum (i.e.~$[{\rm data}] - [{\rm
    fit}]$), normalized to the 1-$\sigma$ errors (faint, horizontal
  solid lines), is shown above the tick-marks.}
\label{fig:fit_all_supp}
\end{center}
\end{figure*}

\begin{figure*}
\begin{center}
\includegraphics[width=0.92\textwidth]{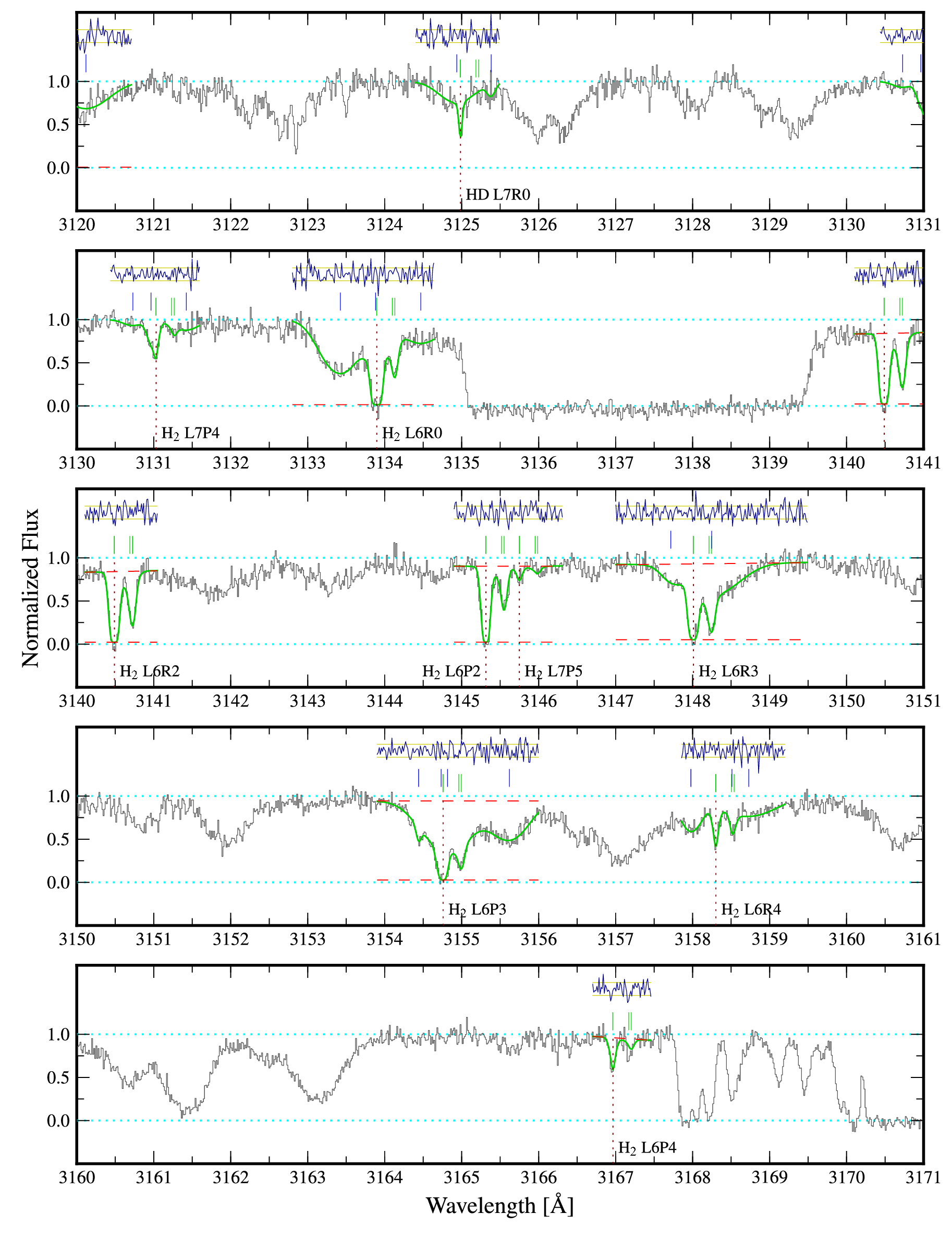}\vspace{-0.5em}
\contcaption{}
\end{center}
\end{figure*}

\begin{figure*}
\begin{center}
\includegraphics[width=0.92\textwidth]{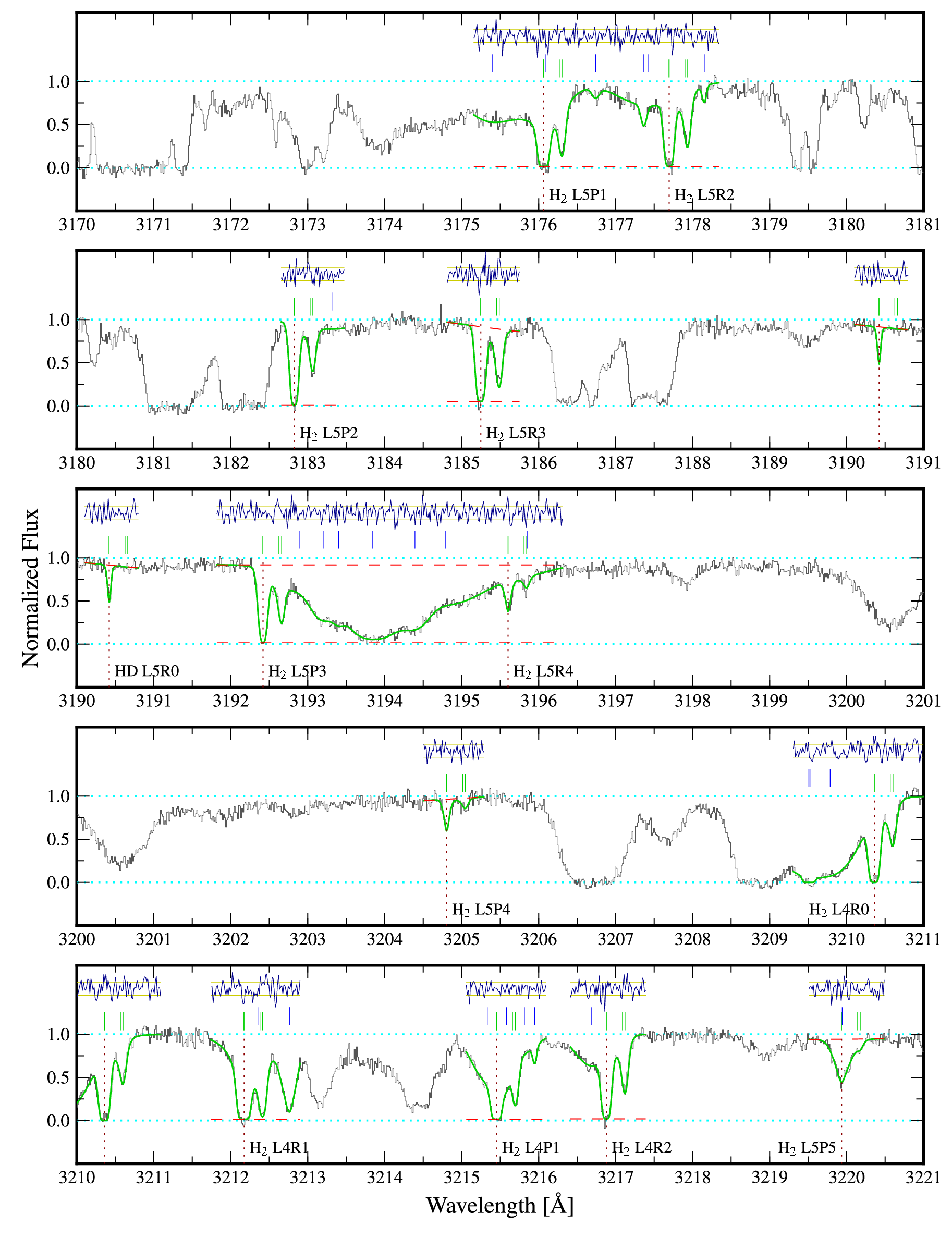}\vspace{-0.5em}
\contcaption{}
\end{center}
\end{figure*}

\begin{figure*}
\begin{center}
\includegraphics[width=0.92\textwidth]{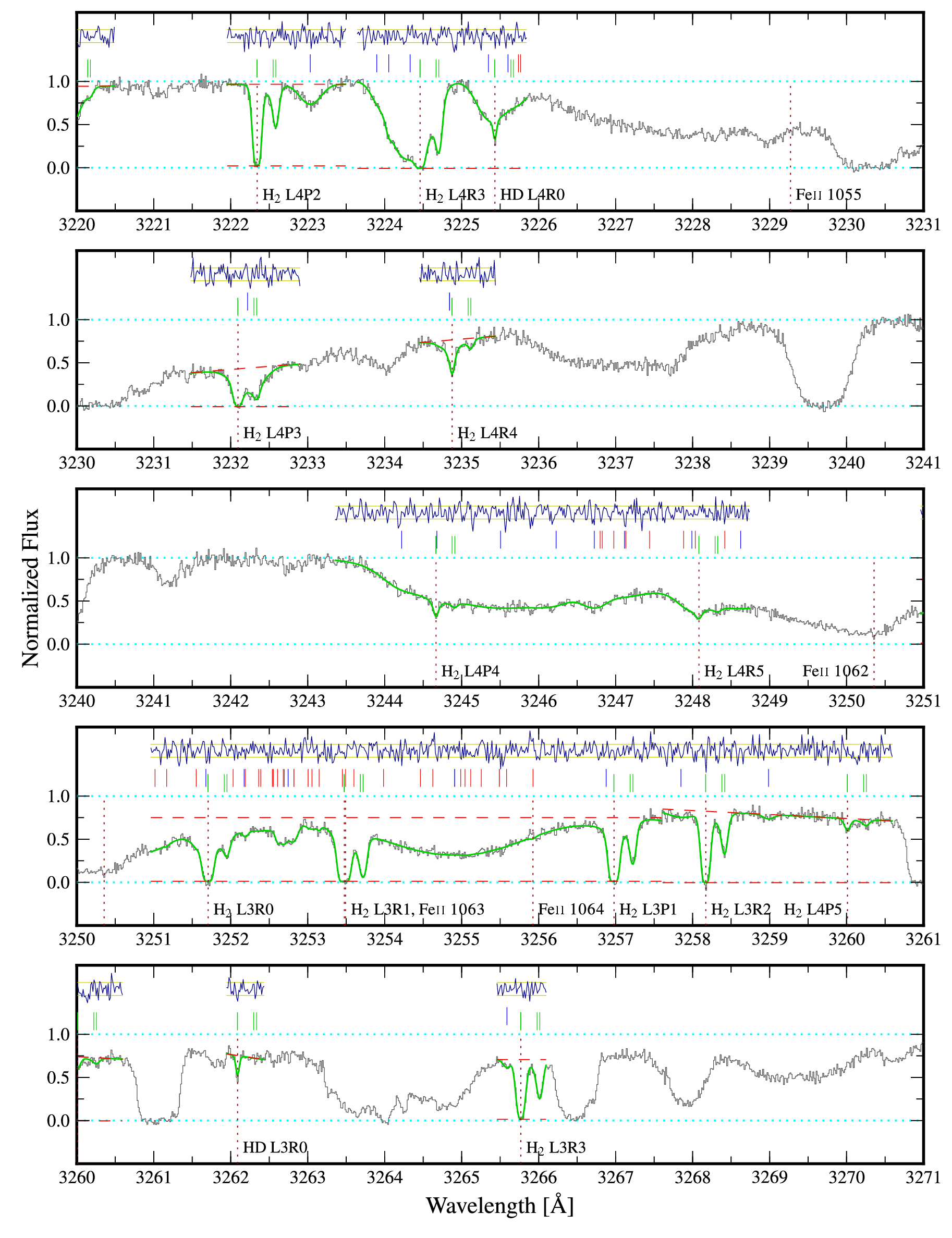}\vspace{-0.5em}
\contcaption{}
\end{center}
\end{figure*}

\begin{figure*}
\begin{center}
\includegraphics[width=0.92\textwidth]{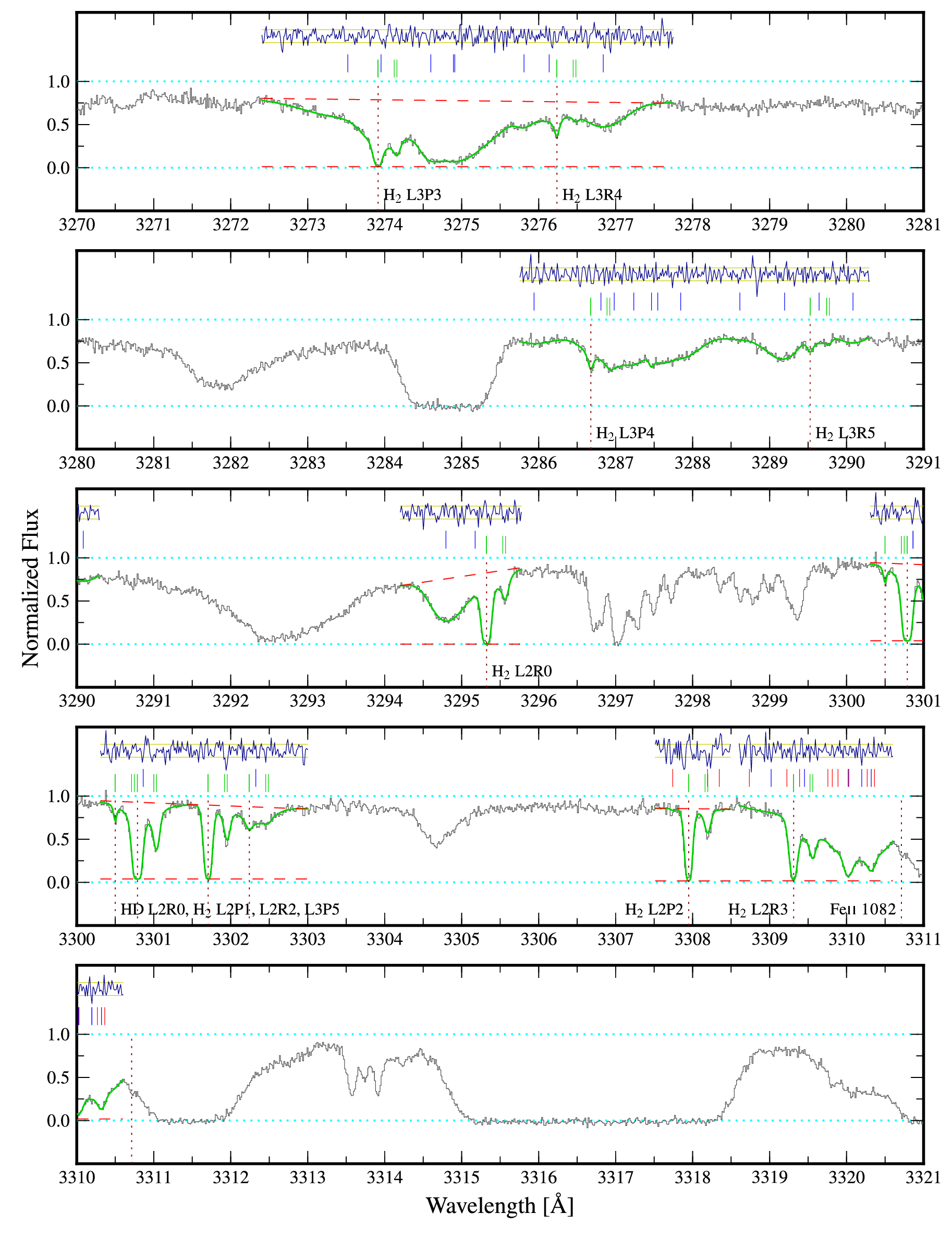}\vspace{-0.5em}
\contcaption{}
\end{center}
\end{figure*}

\begin{figure*}
\begin{center}
\includegraphics[width=0.92\textwidth]{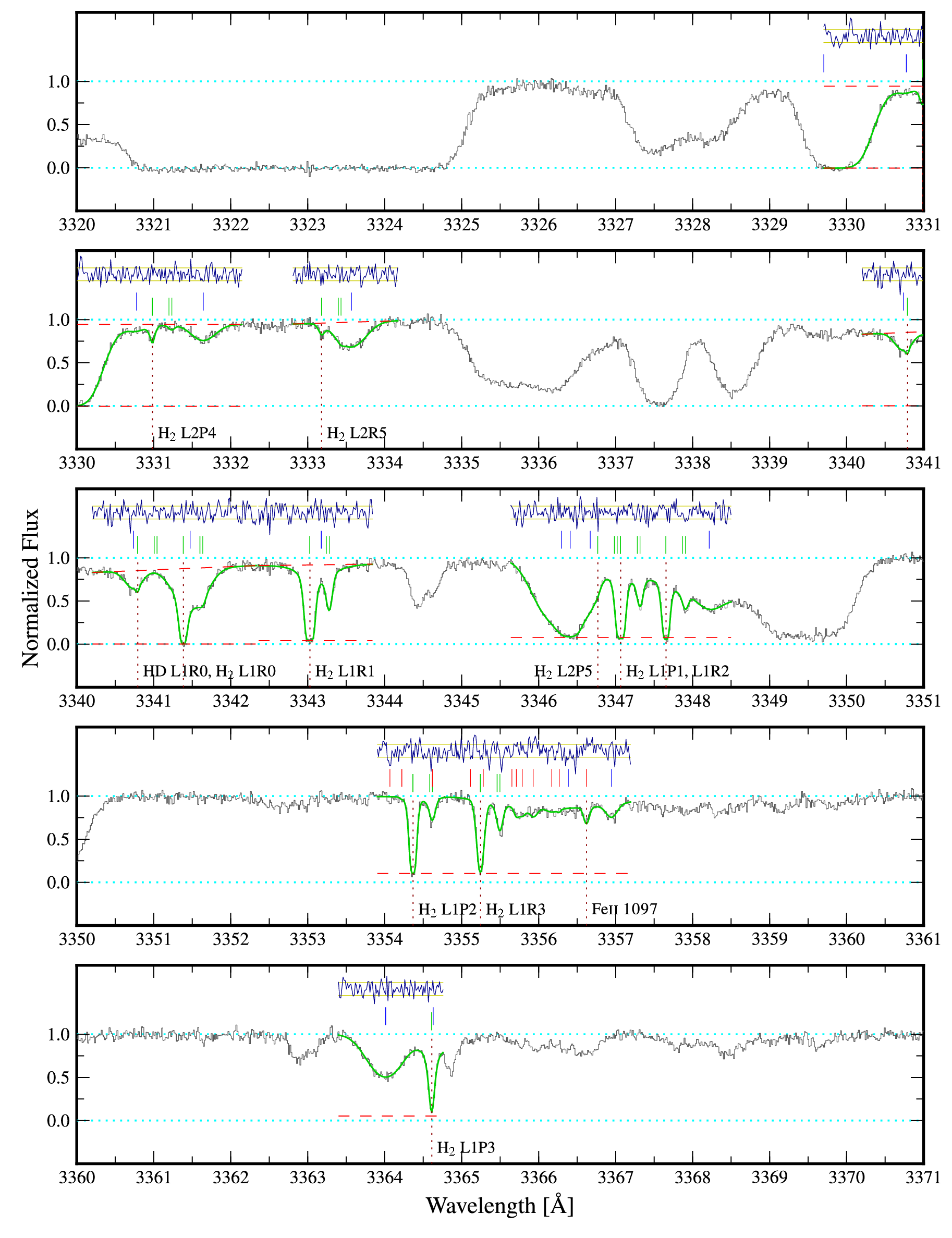}\vspace{-0.5em}
\contcaption{}
\end{center}
\end{figure*}

\begin{figure*}
\begin{center}
\includegraphics[width=0.92\textwidth]{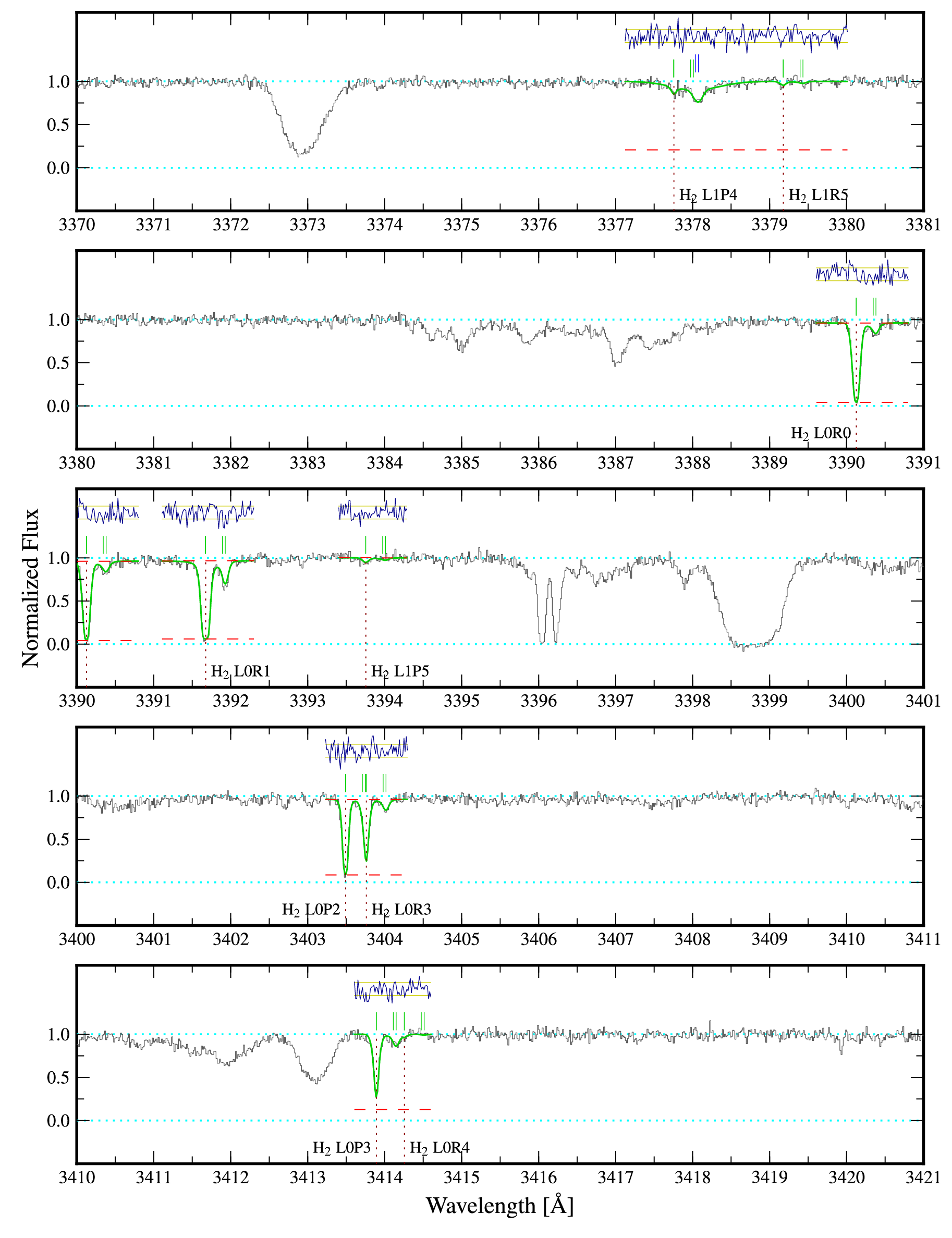}\vspace{-0.5em}
\contcaption{}
\end{center}
\end{figure*}

\begin{figure*}
\begin{center}
\includegraphics[width=0.92\textwidth]{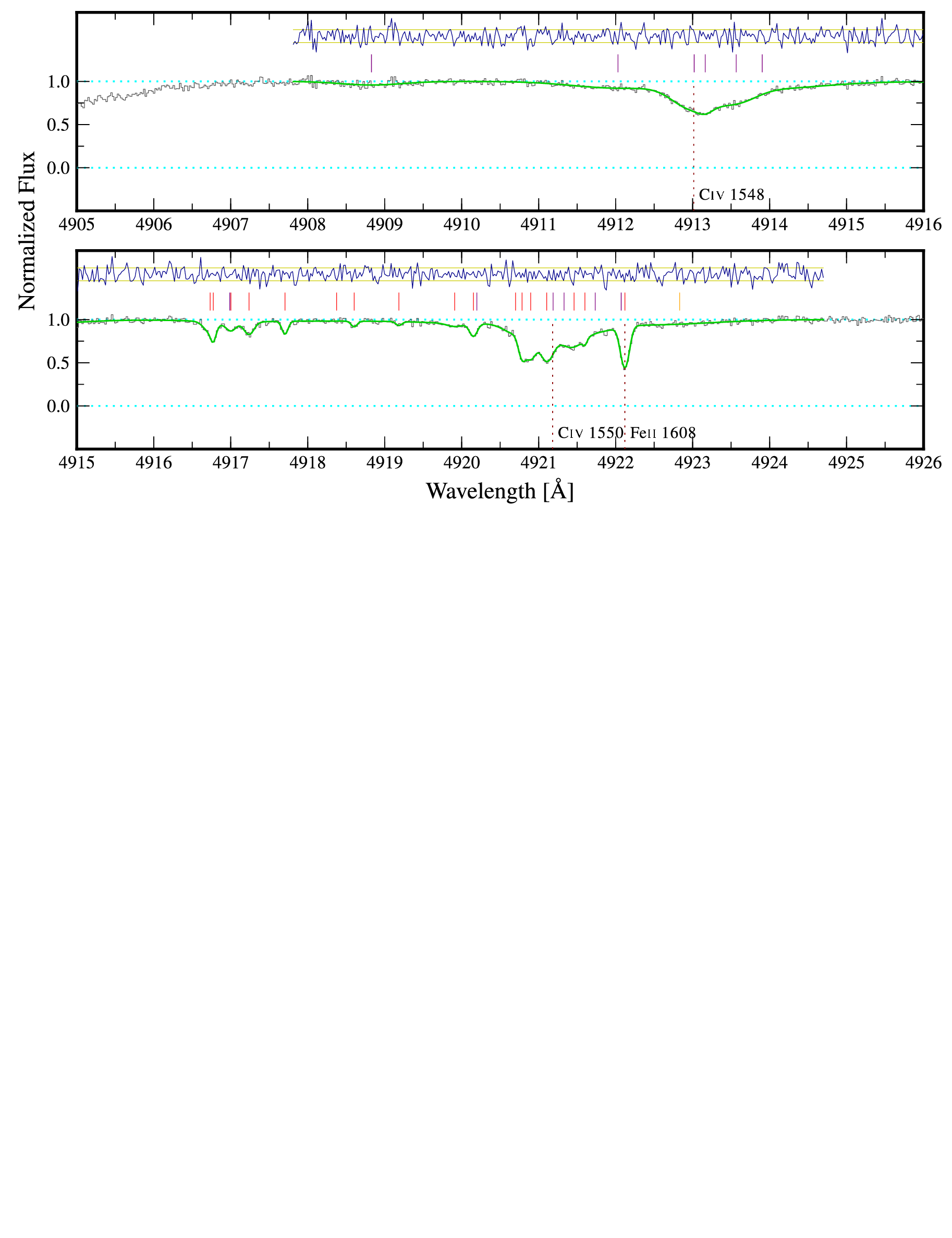}\vspace{-0.5em}
\contcaption{}
\end{center}
\end{figure*}

\end{document}